\documentclass[longbibliography,aps, prx, amsmath, amssymb, amsfonts, twocolumn,superscriptaddress]{revtex4-2}
\usepackage[german,american]{babel}
\usepackage{graphicx}
\usepackage{dcolumn}
\usepackage{bm}
\usepackage{amssymb}
\usepackage{amsmath}
\usepackage{amsfonts}
\usepackage{mathbbol}
\usepackage{latexsym}
\usepackage{dcolumn}
\usepackage{hyperref}
\usepackage{float}
\usepackage[normalem]{ulem}
\usepackage[usenames, dvipsnames]{color}
\usepackage{epstopdf}
\usepackage{comment}
\usepackage[braket,qm]{qcircuit}
\usepackage[ruled]{algorithm2e}
\usepackage{subfigure}

\hypersetup{                    
	colorlinks,
	citecolor=blue,
	filecolor=blue,
	linkcolor=blue,
	urlcolor=blue
}
\usepackage[utf8]{inputenc}

\begin{document}
	\title{Solution of SAT Problems with the Adaptive-Bias Quantum Approximate Optimization Algorithm}
	
	\author{Yunlong Yu}
	\affiliation{ State Key Laboratory of Low Dimensional Quantum Physics, Department of Physics,\\Tsinghua University, Beijing 100084, China}
	\affiliation{Kavli Institute for Theoretical Sciences, University of Chinese Academy of Sciences, Beijing, China}
	
	\author{Chenfeng Cao}
	\affiliation{Department of Physics, The Hong Kong University of Science and Technology,\\ Clear Water Bay, Kowloon, Hong Kong, China}
	
	\author{Xiang-Bin Wang}
	\affiliation{ State Key Laboratory of Low Dimensional Quantum Physics, Department of Physics,\\Tsinghua University, Beijing 100084, China}
	
	\author{Nic Shannon}
	\affiliation{Theory of Quantum Matter Unit, 
		Okinawa Institute of Science and Technology Graduate University, Onna-son, Okinawa 904-0412, Japan}
	
	\author{Robert Joynt}
	\affiliation{Department of Physics, University of Wisconsin–Madison, 1150 University Avenue, Madison, Wisconsin 53706, USA}
	\affiliation{Kavli Institute for Theoretical Sciences, University of Chinese Academy of Sciences, Beijing, China}

	\begin{abstract}
		The quantum approximate optimization algorithm (QAOA) is a promising method for solving certain classical combinatorial optimization problems on near-term quantum devices. When employing the QAOA to $3$-SAT and Max-$3$-SAT problems, the quantum cost exhibits an easy-hard-easy or easy-hard pattern respectively as the clause density is changed. The quantum resources needed in the hard-region problems are out of reach for current NISQ devices. We show by numerical simulations with up to 14 variables and analytical arguments that the adaptive-bias QAOA (ab-QAOA) greatly improves performance in the hard region of the $3$-SAT problems and hard region of the Max-$3$-SAT problems. For similar accuracy, on average, ab-QAOA needs 
     $3$ levels for $10$-variable $3$-SAT problems as compared to $22$ for QAOA.  For $10$-variable Max-$3$-SAT problems, the numbers are $7$ levels and $62$ levels. The improvement comes from a more targeted and more limited generation of entanglement during the evolution. We demonstrate that classical optimization is not strictly necessary in the ab-QAOA since local fields are used to guide the evolution.  This leads us to propose an optimization-free ab-QAOA that can solve the hard-region 3-SAT and Max-3-SAT problems effectively with significantly fewer quantum gates as compared to the original ab-QAOA. 
		Our work paves the way for realizing quantum advantages for optimization problems on NISQ devices.
	\end{abstract}
	\date{\today}
	\maketitle
	
	\section{Introduction}
	
	We are in the Noisy Intermediate-Scale Quantum (NISQ) era for quantum computing~\cite{preskill2018quantum}. NISQ devices such as Sycamore~\cite{sycamore} and Zuchongzhi~\cite{Zuchongzhi} have demonstrated a quantum advantage on the random circuit sampling problem, but this problem is far from practical applications.  Quantum optimization algorithms, such as the quantum adiabatic algorithm(QAA)~\cite{qaa_review,sta_review,bias_qaa,qaa_steering,qaa_catalyst} and the quantum approximate optimization algorithm (QAOA)~\cite{qaoa_farhi,qaoa_lukin}, or the variational quantum algorithm (VQA)~\cite{vqe,cerezo2021variational,bharti2021noisy,theory_vqe,extrapolation, Cao2022quantumvariational}  would have much wider impact, if they could also demonstrate some quantum advantage.  There is hope for this in cases where one has a VQA, in which an outer classical optimizer is employed to train a sequence of parameterized quantum circuits.  If each such circuit has relatively low depth, then noise may be minimized.

    The QAOA aims to solve combinatorial optimization problems and even the lowest depth version has the potential to establish quantum advantages~\cite{farhi2016quantum}. The QAOA is a generalization of the QAA~\cite{qaoa_farhi}, in which the schedule of adiabatic evolution can be modified to produce optimal results.
    For the standard QAOA, it has been experimentally implemented~\cite{sycamore_qaoa,Rydberg_qaoa3,Rydberg_qaoa,Rydberg_qaoa2}.
    There are encouraging results of the QAOA on the MaxCut problems~\cite{qaoa_farhi,analytical_qaoa,qaoa_level23,crooks2018performance,qaoa_lukin,sycamore_qaoa,qaoa_nn}. 
    Simulations of the QAOA give solutions for MaxCut problems on sizes up to $20$ vertices~\cite{qaoa_lukin} with the standard method and up to $54$ qubits with a neural network based method~\cite{qaoa_nn}.  There have been experimental demonstrations in superconducting systems~\cite{sycamore_qaoa} for $23$-qubit graph MaxCut problems. 
    The QAOA seems to be effective for MaxCut problems in the sense of greatly improving on naive adiabatic algorithms. However, there is no evidence to date of a speedup over classical algorithms.
    In this paper we do not attempt to give such evidence.  Rather we seek to improve the QAOA to make it more competitive.  
 
     For the MaxCut problem we have shown a computational speedup over the QAOA when adaptive bias fields are introduced, a modification called the ab-QAOA~\cite{ab_qaoa}. In this paper we pinpoint a problem where QAOA appears to have difficulties, and show that the ab-QAOA greatly outperforms the QAOA. We use the resulting numerical data to pinpoint the strengths of the ab-QAOA.
    
    Specifically, the QAOA encounters difficulties when applied to random $3$-SAT problems~\cite{1_k_sat_qaoa} and random Max-$3$-SAT problems~\cite{reachabiilty_deficits_sat}. 
    These problems have been intensely studied in the framework of classical algorithms ~\cite{mezard2009information,transition_3sat,transition_3sat2,transition_max3sat,garey1979computers,wmaxsat}.  Much is known about their complexity.  A SAT problem or its optimization version Max-SAT problem is defined in terms of $n$ Boolean variables and $m$ clauses. 
    In the classical computing context, the cost for a given accuracy varies with the clause density $\alpha=m/n$.  As $\alpha$ increases, there is an easy-hard-easy pattern in the random $3$-SAT problems and an easy-hard pattern in the random Max-$3$-SAT problems~\cite{transition_3sat,transition_3sat2,transition_max3sat}. Even for the $10$-variable $3$-SAT problems~\cite{1_k_sat_qaoa} and $6$-variable Max-$3$-SAT problems~\cite{reachabiilty_deficits_sat}, the same pattern is evident in the QAOA: in the hard regions, a very large number of levels is required to obtain an accurate ground state. This means that such problems are out of reach of the QAOA on current NISQ devices~\cite{sycamore_qaoa}, and prospects are dim for the near future. It is intriguing that the same easy-hard patterns are evident in both the QAOA and the classical algorithms, and that, as we will show, the patterns are nearly absent in the ab-QAOA. 
	
In the hard-region Max-$3$-SAT problems, this lack of convergence in the QAOA is known as the reachability deficits~\cite{reachabiilty_deficits_sat,reachability_deficits_graph}. In physics terms, one can track this back to a large amount of frustration in the Ising variables.  This happens locally when a triangle of spins has antiferromagnetic interactions and similar problems repeat on multi-variable sets~\cite{wannier1950antiferromagnetism}. The overwhelming overhead of QAOA in the hard-region Max-$3$-SAT problems, \textit{i.e.} the reachability deficits, is not strictly related to barren plateaus~\cite{barren_plateaus,barren_plateaus_exp,gradientfree_barrenplateau,barren_plateaus_ml,barren_plateaus_ent,barren_plateaus_ent2,barren_plateaus_avoid,barren_plateaus_noise}, defined to be when the variance of the gradients vanishes exponentially with the system size $n$, making the cost function hard to train.
    Reachability deficits can occur even in the absence of the barren plateaus, \textit{e.g.} for large clause density $\alpha$ with a fixed $n$. Nevertheless, the hard-region SAT problems do seem to exhibit a small variance of the energy gradients~\cite{1_k_sat_qaoa}. 

	There is intense research activity to further improve the performance of QAOA.  This includes heuristic initialization strategies~\cite{qaoa_lukin,tqa_qaoa}, modifications of the mixing Hamiltonian~\cite{ab_qaoa,general_qaoa,adapt_qaoa}, adjusting the cost function~\cite{rqaoa,cvarqaoa}, the warm-start strategy~\cite{qaoa_warmstart}, utilizing adiabaticity~\cite{cdqaoa,cdqaoa2} and using machine learning~\cite{traingqaoa}. However, little is known about their performances on the easy-hard-easy or easy-hard transitions on the relevant SAT problems. 
	
    In the ab-QAOA, longitudinal adaptive bias fields are incorporated into the mixing Hamiltonian, which are updated based on the expectation values of the Pauli $Z$ operators~\cite{ab_qaoa}. The ab-QAOA is a generalization of the QAOA and, as stated above, a substantial and scalable speedup over the QAOA on the MaxCut problem has been observed. Furthermore, unlike most other adaptive QAOA variants~\cite{adapt_qaoa,ADAPTqaoa_ent}, the ab-QAOA requires no more measurements than the QAOA. 
	
    The three main outcomes of this paper are as follows.
    \begin{enumerate}
        \item We show that the ab-QAOA improves over the QAOA for certain SAT problems with easy-hard-easy or easy-hard patterns where QAOA does not perform well. Our strategy is to first demonstrate the speedup of the ab-QAOA over the QAOA for the relevant SAT problems.
        \item We analyze the characteristics of the results and increase our understanding of the reasons for the improved performance of the ab-QAOA.
        \item We propose an optimization-free ab-QAOA to reduce the overhead of gradient calculations and show that the easy-hard-easy and easy-hard transitions are not evident in this optimization-free version.
    \end{enumerate}
    The fact that the ab-QAOA is less subject to the well-known easy-hard-easy or easy-hard transitions is the evidence that it is \textit{qualitatively} superior to the QAOA.  Taken together, these features mean that the ab-QAOA is a considerable step forward.  
    None the less these improvements on the QAOA do not imply by themselves that our algorithm produces a speedup of classical algorithms for SAT problems. 
    Establishing quantum advantages in this context would require a separate analysis, which lies beyond the scope of the present work.

	The paper is organized as follows. In Sec.~\ref{sec:ab_qaoa}, we will give a detailed description of the QAOA and ab-QAOA, including some modifications of the ab-QAOA relative to the version in Ref.~\cite{ab_qaoa}. In Sec.~\ref{sec:sat}, a discussion of the relevant details of the special version of $3$-SAT or Max-$3$-SAT problems considered in this work , the $1$-$3$-$\mathrm{SAT}^+$ or Max-$1$-$3$-$\mathrm{SAT}^+$ problems and the easy-hard-easy or easy-hard patterns can be found. In Sec.~\ref{sec:result}, 
	the relative performances of QAOA and ab-QAOA on the $1$-$3$-$\mathrm{SAT}^+$ and Max-$1$-$3$-$\mathrm{SAT}^+$ problems are given.  
	In Sec.~\ref{sec:analysis} we analyze the advantages of the ab-QAOA: its targeted nature of the entanglement in the evolution, which is related to many-body localization and the increased adiabaticity in the discrete time evolution. In Sec.~\ref{sec:of_abqaoa}, we demonstrate that an optimization-free version of the ab-QAOA can solve the hard-region $1$-$3$-$\mathrm{SAT}^+$ or Max-$1$-$3$-$\mathrm{SAT}^+$ problems effectively and with much fewer quantum resources. Our conclusions are summarized in Sec.~\ref{sec:conclusion}.
	
	\section{ Adaptive-bias quantum approximate optimization algorithm}\label{sec:ab_qaoa}
	The standard QAOA is a quantum-classical hybrid algorithm to solve the combinatorial problems~\cite{qaoa_farhi}. 
	The problem is encoded in the $n$-qubit cost Hamiltonian $H_\mathrm{C}$,  whose ground state is the desired solution. In the cases investigated to date, $H_\mathrm{C}$ is a classical Ising model that only contains Pauli $Z$ operators~\cite{np_ising}. The quantum part of the standard QAOA starts from $|\psi_0^\mathrm{s}\rangle$, the ground state of the mixing Hamiltonian $H_\mathrm{M}^\mathrm{s}=\sum_j X_j$, where $X_j$ is the Pauli $X$ operator acting on $j^\mathrm{th}$ qubit. The unitary operators $\exp(-i\beta_k H_\mathrm{M}^\mathrm{s})$ and $\exp(-i\gamma_k H_\mathrm{C})$ are alternately applied to $|\psi_{0}^\mathrm{s}\rangle$ $p$ times, where $p$ is the level. The output state of the QAOA is,
	\begin{align}
		|\psi_\mathrm{f}^{\mathrm{s}}(\vec{\gamma},\vec{\beta}) \rangle=
		\prod_{k=1}^{p}
		\mathrm{e}^{-i\beta_{k}H_\mathrm{M}^\mathrm{s}} \mathrm{e}^{-i\gamma_{k}H_\mathrm{C}}|\psi_{0}^{\mathrm{s}}\rangle. \label{eq:qaoa}
	\end{align}
	Here we use the vector expression $\vec{\gamma},\vec{\beta}$ to represent a set of parameters $\{\gamma_1,\cdots,\gamma_p\}$ and $\{\beta_1,\cdots,\beta_p\}$. The operators with subscript $k$ are always on the left of those with $k-1$. The classical part of the QAOA is the iterative optimization of $\vec{\gamma}$ and $\vec{\beta}$ according to the measurement of $\langle H_\mathrm{C} \rangle$, the expectation value of $H_\mathrm{C}$ in $|\psi_\mathrm{f}^\mathrm{s}(\vec{\gamma},\vec{\beta})\rangle$. Note that whether $|\psi_{0}^\mathrm{s}\rangle$ is taken to be $|-\rangle^{\otimes n}$ or $|+\rangle^{\otimes n}$ has no effect on the classical optimization procedure, since $(\vec{\gamma},\vec{\beta})$ with $|-\rangle^{\otimes n}$ yields the same $\langle H_\mathrm{C} \rangle$ as $(\vec{\gamma},-\vec{\beta})$ with $|+\rangle^{\otimes n}$~\cite{ab_qaoa}.

	\begin{figure*}[ht]
		\centering
		\includegraphics[scale=0.4]{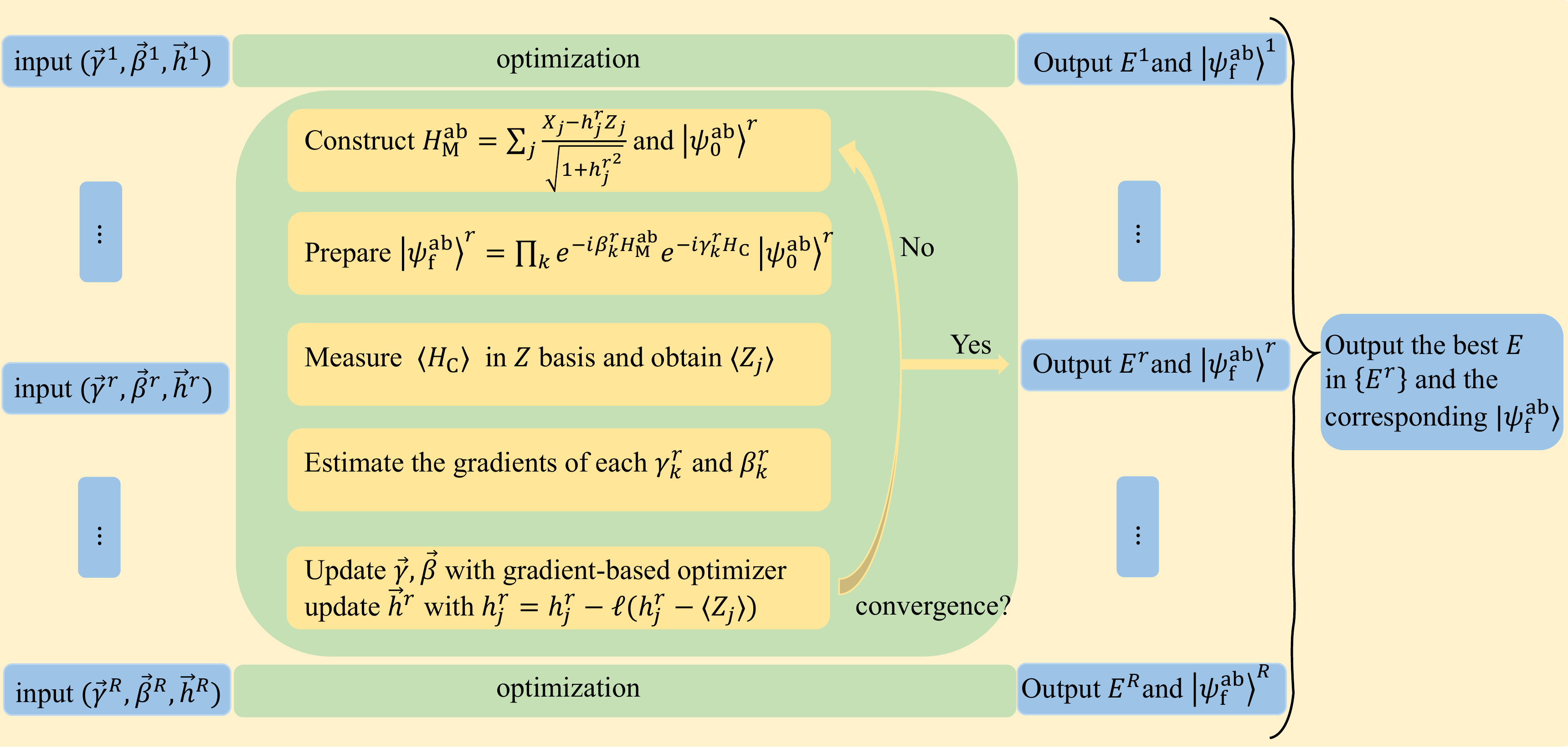}
		\caption{A schematic of the optimization procedure of the adaptive-bias quantum approximate optimization algorithms (ab-QAOA) in level $p$. The optimization of $R$ samples can be done in parallel and  the lowest energy among $R$ optimized results and the corresponding state $|\psi_\mathrm{f}^\mathrm{ab}\rangle$ are the outputs in this level. For a single sample $r$, in each optimization iteration, the updates of $\vec{\gamma}$ and $\vec{\beta}$ are different from $\vec{h}$. To determine the direction of $\gamma_k$ or $\beta_k$ in next iteration, additional preparations and measurements of $|\psi_\mathrm{f}^{\mathrm{ab}}\rangle$  with slightly moved $\gamma_{k}$ or $\beta_k$ are needed. However, this overhead is not necessary for $h_j$, whose direction is determined by $\langle Z_j \rangle$. This schematic also applies to the QAOA if we set $h_j=0$ and $\ell=0$. }
		\label{fig:abqaoa}
	\end{figure*}
 
	The recently proposed ab-QAOA is a generalization of the QAOA~\cite{ab_qaoa}. Local longitudinal bias fields $\vec{h}=\{h_1,h_2,...,h_n\}$ are incorporated into the mixing Hamiltonian, giving,
	\begin{align}
		H_\mathrm{M}^\mathrm{ab}(\vec{h})=\sum_j \frac{X_j-h_j Z_j}{\sqrt{1+h_j^2}},
		\label{eq:ab_mixing}
	\end{align}
	where $Z_j$ is the Pauli $Z$ operator acting on the $j^\mathrm{th}$ qubit. The choice of the $h_j$ is discussed below.  The starting state $|\psi_{0}^\mathrm{ab}\rangle$ is always reinitialized to be the ground state of $H_\mathrm{M}^\mathrm{ab}$.  This is the key feature of the ab-QAOA that gives its advantages over the QAOA, as will be discussed in Sec.~\ref{sec:analysis}. Thus the output state of the ab-QAOA is,
	\begin{align}
		|\psi_\mathrm{f}^{\mathrm{ab}}(\vec{\gamma},\vec{\beta},\vec{h}) \rangle=
		\prod_{k=1}^{p}
		\mathrm{e}^{-i\beta_{k}H_\mathrm{M}^\mathrm{ab}(\vec{h})} \mathrm{e}^{-i\gamma_{k}H_\mathrm{C}}|\psi_{0}^{\mathrm{ab}}(\vec{h})\rangle.\label{eq:ab_qaoa}
	\end{align}

	The $n$ extra bias fields parameters $\vec{h}$ are not optimized, but rather updated according to the prescription $h_j\rightarrow h_j-\ell(h_j-\langle \psi_\mathrm{f}^{\mathrm{ab}}|Z_j|\psi_\mathrm{f}^{\mathrm{ab}}\rangle )$ in each iteration. $\ell$ is the learning rate and we take to be $\ell=0.4$ in this paper. $\vec{\gamma}$ and $\vec{\beta}$ are optimized in the usual way based on $\langle H_\mathrm{C}\rangle$. Since the energy measurement is in the $Z$ basis, $\langle \psi_\mathrm{f}^{\mathrm{ab}}|Z_j|\psi_\mathrm{f}^{\mathrm{ab}}\rangle $ can be obtained without any additional overhead. When $h_j\rightarrow 0$ and $\ell \rightarrow 0$, the ab-QAOA is equivalent to the QAOA. A schematic the of ab-QAOA algorithm is shown in Fig.~\ref{fig:abqaoa}.
    
	In Eq.~\eqref{eq:ab_mixing}, the Schmidt norm of the operator on the $j^\mathrm{th}$ qubit is normalized to identity (it squares to the identity operator). This has the advantage that each $\beta_{k}$ can be restricted to the interval $[0,\pi]$, unlike in Ref.~\cite{ab_qaoa}. On the $j^\mathrm{th}$ qubit, we define the rotation angle $d_j$ with the relationships,
	\begin{equation}
	   \cos d_j=\frac{1}{\sqrt{1+h_j^2}},\quad
	   \sin d_j=\frac{h_j}{\sqrt{1+h_j^2}},
	\end{equation}
	and the rotation operator around the $\hat{y}$ axis is $R_y^j(d_j)=\exp(-i d_j Y_j/2)$, where $Y_j$ is the Pauli $Y$ operator on qubit $j$. Eq.~\eqref{eq:ab_mixing} can then be rewritten as, 
	\begin{equation}
		\begin{split}
			H_\mathrm{M}^\mathrm{ab}(\vec{h})&=\sum_j R_y^j(d_j) X_j R_y^{j\dagger}(d_j),\\
			&=\tilde{R}_y(\vec{d}\,) H_\mathrm{M}^\mathrm{s} \tilde{R}_y^\dagger(\vec{d}\,),
		\end{split}
	\end{equation}
	where $\tilde{R}_y(\vec{d}\,)=\bigotimes_j  R_y^j(d_j)$. The $[0,\pi]$ period of $\beta_{k}$~\cite{qaoa_farhi} is not affected by the update of the bias fields. The eigenvalues of $H_\mathrm{C}$ are integers for the models considered in this paper.  Hence $\gamma_k$ can be restricted to $[0,2\pi]$. The ab-QAOA starting state $|\psi_{0}^\mathrm{ab}\rangle$ can be obtained by rotating $|0\rangle^{\otimes n}$ around $\hat{y}$ axis in contrast with applying Hadamard gates to $|0\rangle^{\otimes n}$ for $|\psi_{0}^{\mathrm{s}}\rangle$ in the QAOA. Besides the initial state preparation, no additional quantum gates are needed to prepare a $p$ level $|\psi_\mathrm{f}^\mathrm{ab}\rangle$ compared with $|\psi_\mathrm{f}^\mathrm{s}\rangle$~\cite{ab_qaoa}, so again there is no additional overhead.
	
	The energy landscapes of the ab-QAOA at high levels are generally complicated. A good initial guess of $(\vec{\gamma},\vec{\beta},\vec{h})$ can help to reduce the searching space and speed up the convergence of the classical optimization. Several heuristic initialization strategies have been proposed for the QAOA~\cite{qaoa_lukin,tqa_qaoa}. In this paper we modify the Trotterized quantum annealing (TQA) initialization strategy~\cite{tqa_qaoa} as discussed below and in Algorithm~\ref{alg:tqa} and the Fourier strategy ~\cite{qaoa_lukin} as discussed in Appendix~\ref{sec:modified_lukin} to make them more efficient.  These methods are then used for both the QAOA and ab-QAOA in order to compare them on an equal basis.

	Combining the ideas of the TQA method in \cite{tqa_qaoa} and the Fourier strategy in \cite{qaoa_lukin}, we propose a modified TQA method. In level $p$, the optimization starts from $R$ points in parallel. Among $R$ optimized results, the lowest energy with the corresponding optimal state are taken to be the outputs in this level as shown in Fig.~\ref{fig:abqaoa}. In the $R$ initial points, the components of all the bias field parameters $\{\Vec{h}^r\}$ are randomly chosen from $\{1,-1\}$. The first set $(\vec{\gamma}^1,\vec{\beta}^1)$ is initialized with the original TQA method and the other sets differ from $(\vec{\gamma}^1,\vec{\beta}^1)$ by a random amount according to the scheme shown in Algorithm~\ref{alg:tqa}. This modified TQA is applied to the ab-QAOA and the QAOA (where the bias fields and the learning rate $\ell$ are initialized to be $0$). The source codes of this modified TQA method are available in~\cite{codes}.

\begin{algorithm}
		\caption{Modified TQA method for ab-QAOA}
		\label{alg:tqa}
		\SetKwInOut{Return}{Return}
		\KwIn{level $p$, total number of samples $R$ }
		\KwOut{$R$ initial points for optimization}
		\For{$r=1$ \KwTo $R$}{
			Randomly generate bias fields parameters $\vec{h}^{r}$ with each components to be $1$ or $-1$.\\
			\eIf{$r\,\mathrm{is}\, 1$}{
				Initialize the components of $\vec{\gamma}^r$ and $\vec{\beta}^r$ according to a linear schedule,
				\begin{equation}
					\begin{split}
						\gamma_k^r&=\frac{k-1}{p}\delta t,\\
						\beta_k^r&=(1-\frac{k-1}{p})\delta t.
					\end{split}
				\end{equation}
			}
			{
				Add some random numbers to the components of $\vec{\gamma}^1$ and $\vec{\beta}^1$, 
				\begin{equation}
					\begin{split}
						\gamma_k^r&=\gamma_k^1+\mathrm{Ran}(\gamma_k^1),\\
						\beta_k^r&=\beta_k^1+\mathrm{Ran}(\beta_k^1),
					\end{split}
				\end{equation}
			}	
		}
		\Return{$R$ initial points $\{(\vec{\gamma}^r,\vec{\beta}^r,\vec{h}^r)\}$.}
	\end{algorithm}

	In Algorithm~\ref{alg:tqa}, the superscript $r$ runs from $1$ to $R$ and labels the different points of the initialization and the subscript $k$ means the $k^\mathrm{th}$ component in the vector. The random number $\mathrm{Ran}(u)$ is a normally-distributed number multiplied by a rescaled factor $\xi$, $\mathrm{Ran}(u)=\xi\mathrm{Norm}(0,u^2)$, where $0$ is the mean value and $u^2$ is the variance. In our calculations, $\delta t=\xi=0.6$ and $R=10$.

	\section{SAT problems} 
	\label{sec:sat}

	A satisfiability (SAT) problem is defined in terms of $n$ Boolean variables $\{x_j\}_{j=1}^n$ taking values from $\{0 \,(\mathrm{False}),1\,(\mathrm{True})\}$ and $m$ clauses $\{C_a\}_{a=1}^m$~\cite{mezard2009information}.  The negation of variable $x_j$ is $\overline{x}_j=1-x_j$. A literal $y_j$ is either a variable or its negation $\overline{x}_j$, \textit{i.e.} $y_j\in\{x_j,\overline{x}_j\}$. A clause $C_a$ can be written as some literals connected by logical OR ($\lor$), for example $C_1=y_1\lor y_2 \lor y_3 $. In the usual SAT problem a clause $C_a$ is satisfied if and only if at least one literal takes value $1$. A SAT problem can be represented by the combination of $m$ clauses connected by logical AND ($\land$),
	\begin{align}
		F=C_1\land C_2\land\cdots\land C_m, \label{eq:cnf}
	\end{align}
	which is called conjunctive normal form (CNF). The conjunctive normal form $F$ is satisfied if and only if all clauses $\{C_a\}_{a=1}^m$ are satisfied. 
 
    The SAT problem is a decision problem, whose goal is to answer the question whether there exits an assignment of $\{x_j\}_{j=1}^n$ such that the formula $F$ is satisfied (SAT) or not (UNSAT). The corresponding optimization version is the Max-SAT problem which aims to find the assignment that violates the smallest number of clauses. Generally, in the Max-SAT problem, each clause can be assigned a weight and the aim of such weighted Max-SAT problem is to find the assignment that minimizes the sum of all the weights in the unsatisfied clauses~\cite{wmaxsat}. We consider a modified version of 3-SAT called the  $1$-$3$-$\mathrm{SAT}^+$ problem, in which each clause contains exactly $3$ positive literals, where the positive literal means a literature $y_j$ only represents the positive variable $x_j$, and a satisfied clause contains exactly one true literal.  This problem is NP-complete in general while its optimization version  Max-$1$-$3$-$\mathrm{SAT}^+$ and weighted Max-$1$-$3$-$\mathrm{SAT}^+$ are NP-hard~\cite{garey1979computers,wmaxsat}. 
	
	Penalty terms are introduced to convert the $1$-$3$-$\mathrm{SAT}^+$ problem or Max-$1$-$3$-$\mathrm{SAT}^+$ problem to an Ising cost Hamiltonian~\cite{qubo,np_ising}. Finding the solution for the original problem is equivalent to finding the ground energy or the ground state of an Ising-type Hamiltonian. Note that the QAOA and ab-QAOA are able to solve both Max-$1$-$3$-$\mathrm{SAT}^+$ and $1$-$3$-$\mathrm{SAT}^+$ problems. In the former problem, we need to find the exact ground state, while in the latter problem, we just need to know whether the ground energy is smaller than a threshold $E_{\mathrm{th}}$ (SAT) or not (UNSAT) \cite{1_k_sat_qaoa}, where the threshold $E_{\mathrm{th}}$ is $0.5$ in this paper. The penalty terms for  $1$-$3$-$\mathrm{SAT}^+$ problems are~\cite{1_k_sat_qaoa,qubo,np_ising},
	\begin{align}
		H_\mathrm{C}=\sum_{a=1}^m (y_{a1}+y_{a2}+y_{a3}-1)^2, \label{eq:penalty}
	\end{align}
	where $y_{aj}$ is the $j^\mathrm{th}$ positive literal in the $a^\mathrm{th}$ clause $C_a$. A satisfied clause with only one true literal contributes $0$ in the penalty terms in Eq.~\eqref{eq:penalty} and an unsatisfied clause contributes $1$ or $4$. The values $1$ and $4$ have little effect in the process of finding solutions since whether the problem is SAT or UNSAT is only determined by whether $H_\mathrm{C}=0$ is satisfied or not.
	
	If we replace each $y_{aj}$ appearing in Eq.~\eqref{eq:penalty} with $(1-Z_{aj})/2$, then the penalty terms in Eq.~\eqref{eq:penalty} can be rewritten as an Ising Hamiltonian,
	\begin{align}
		H_\mathrm{C}=\frac{1}{4}\sum_{a=1}^{m} (Z_{a1}+Z_{a2}+Z_{a3}-1)^2, \label{eq:costH} 
	\end{align}
	where the solution for the $1$-$3$-$\mathrm{SAT}^+$ problem is encoded in the ground energy of Eq.~\eqref{eq:costH}. 
	This means the variables taking value $0$ (False) and $1$ (True) are represented by the eigenstates $|0\rangle$ (eigenvalue $1$) and $|1\rangle$ (eigenvalue $-1$) of $Z_{aj}$ respectively. The eigenvalues of Eq.~\eqref{eq:costH} are always integers, so $\gamma_{k}$ can be restricted to the interval $[0,2\pi]$ for both the QAOA and the ab-QAOA. We will refer to $H_\mathrm{C}$ as the cost Hamiltonian in the following.
 
\begin{figure}[htb]
		\centering
		\includegraphics[scale=0.55]{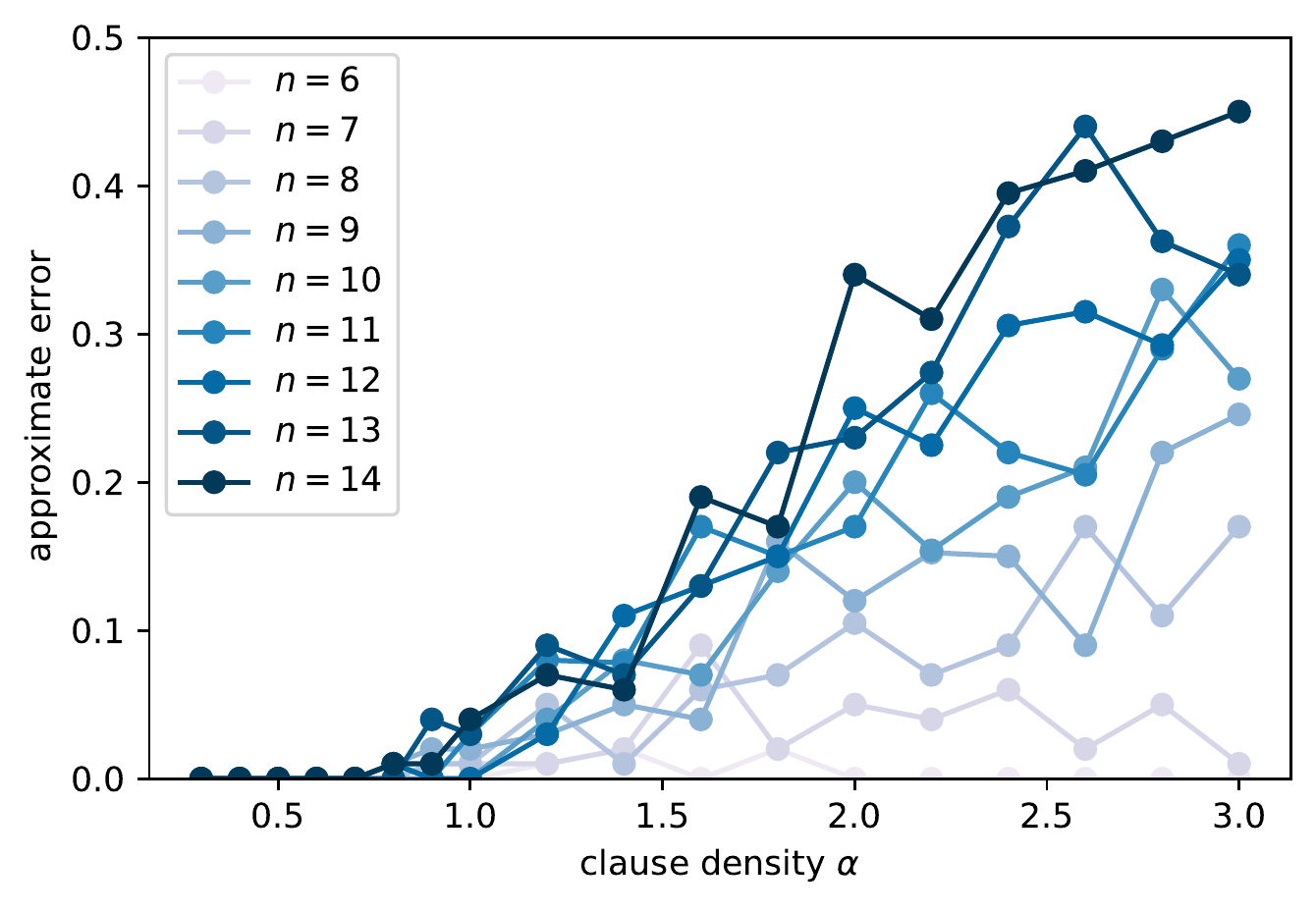}
		\caption{The approximation error (number of violated clauses in the ground state of Eq.~\eqref{eq:costH} minus that in the real solution) as a function of clause density $\alpha$ for different $n$. Each point is the average over $100$ different realizations.}\label{fig:error}
	\end{figure}

    The ground state of Eq.~\eqref{eq:costH} is not necessarily the exact solution to the corresponding Max-$1$-$3$-$\mathrm{SAT}^+$ problem. When $\alpha$ is small, it is an exact one in contrast to an approximate one when $\alpha$ is large. As analyzed in Ref.~\cite{1_k_sat_qaoa} and shown in Fig.~\ref{fig:error}, the approximation error (number of the violated clauses in the ground state of Eq.~\eqref{eq:costH} minus that in the real solution) is within $1$, so the Ising Hamiltonian in Eq.~\eqref{eq:costH} can be used as a good approximation to the Max-$1$-$3$-$\mathrm{SAT}^+$ problem Hamiltonian. 
    
    An accurate mapping to the Ising Hamiltonian can be obtained by reducing the $3$-SAT problem with $m$ clauses to a maximal independent set (MIS) problem on the  graph with $O(m)$ vertices~\cite{np_ising}. However, when $m$ is large, this is obviously out of reach for NISQ devices. In fact, the ground state of Eq.~\eqref{eq:costH}
    encodes the solution for a weighted Max-$1$-$3$-$\mathrm{SAT}^+$ problem with literal-dependent weights $\{w_a(y_{a1},{y_{a2}},y_{a3})\}$. When the literals $y_{a1},{y_{a2}},y_{a3}$ in the clause $C_a$ are all true, $w_a=4$ and $w_a=1$ otherwise.  We emphasize that the Hamiltonian in Eq~\eqref{eq:costH} can be regarded as either an approximation to the Max-$1$-$3$-$\mathrm{SAT}^+$ problem, similar to Ref.~\cite{1_k_sat_qaoa}, or an exact description of the weighted Max-$1$-$3$-$\mathrm{SAT}^+$ problem. We do not distinguish them and call Eq.~\eqref{eq:costH} the Hamiltonians of the Max-$1$-$3$-$\mathrm{SAT}^+$ problems henceforth.  
	
	Let there be $N_\mathrm{SAT}$ SAT instances among $N$ problem instances.  The SAT probability in the SAT problems is defined as,
	\begin{align}
		P_\mathrm{SAT}=\frac{N_\mathrm{SAT}}{N}.
	\end{align} 
	The key parameter in the $3$-SAT problems we consider is the clause density $\alpha$~\cite{mezard2009information,transition_3sat,transition_3sat2,transition_max3sat}.  This is defined as $\alpha=m/n$. There is a SAT-UNSAT phase transition from $P_\mathrm{SAT}=1$ to $P_\mathrm{SAT}=0$ in the random $3$-SAT problems, across a critical clause density $\alpha_\mathrm{c}$. The critical clause density $\alpha_\mathrm{c}$ of the $1$-$3$-$\mathrm{SAT}^+$ problems is the region $\alpha_\mathrm{c}\in (0.546,0.644) $~\cite{transition_ec3}. The classical computational cost suffers from an easy-hard-easy pattern, where the problems near $\alpha_\mathrm{c}$ are known to be the hardest~\cite{mezard2009information,transition_max3sat}. It was found in~\cite{1_k_sat_qaoa} that the QAOA also follows an easy-hard-easy pattern in the same region. 
	
	For the Max-$3$-SAT problems, there is an easy-hard pattern in both classical cost~\cite{transition_max3sat} and quantum cost (QAOA levels)~\cite{reachabiilty_deficits_sat}. The Max-$3$-SAT problems in the hard region solved by QAOA exhibit the reachability-deficit phenomenon~\cite{reachabiilty_deficits_sat}, which means that a large number of levels are required to obtain the ground state. A detailed theoretical analysis of this region does not yet exist, but barren plateaus appear in the hard region \cite{1_k_sat_qaoa}. 
	
	\section{Numerical simulations}
	\label{sec:result}
	
	In this section we compare numerically the performance of QAOA and ab-QAOA when applied to the $1$-$3$-$\mathrm{SAT}^+$ problems and Max-$1$-$3$-$\mathrm{SAT}^+$ problems. The problem instances are randomly generated $1$-$3$-$\mathrm{SAT}^+$ or Max-$1$-$3$-$\mathrm{SAT}^+$ problems with $6\sim14$ variables ($6\sim14$ qubits for Eq.~\eqref{eq:costH}) and different clause densities. The raw data of the problem definition can be found in~\cite{codes}. $6\sim14$ variables is not a large number but the necessity of doing many realizations of the disorder for each value of $\alpha$ and $p$ limits the size of the system.  The QAOA or ab-QAOA is initialized according to the modified TQA method mentioned in Sec.~\ref{sec:ab_qaoa}. The level needed for a fixed accuracy of QAOA and ab-QAOA is used as a criterion for the quantum cost, since the number of quantum gates in a single level is the same for both QAOA and ab-QAOA~\cite{ab_qaoa}. For the $10$-variable  $1$-$3$-$\mathrm{SAT}^+$ problem, we take $\alpha_\mathrm{c}\approx 0.6$. 
	
	\subsection{Success probability} \label{sec:result_energy}
	
	For the decision $1$-$3$-$\mathrm{SAT}^+$ problems, the success probability is used for comparing the relative performances of the QAOA and ab-QAOA. For a given problem instance solved by the QAOA or ab-QAOA, if it is actually SAT and the QAOA or ab-QAOA gives the answer SAT or if it is actually UNSAT and the QAOA or ab-QAOA gives the answer UNSAT, this is called a QAOA (or ab-QAOA) successful instance. We define the success probability as,
	\begin{align}
		P_\mathrm{succ}=\frac{N_\mathrm{succ}}{N},
	\end{align} 
	where $N_\mathrm{succ}$ is the number of successful instances.

\begin{figure}[htb]
		\centering
		\includegraphics[scale=0.55]{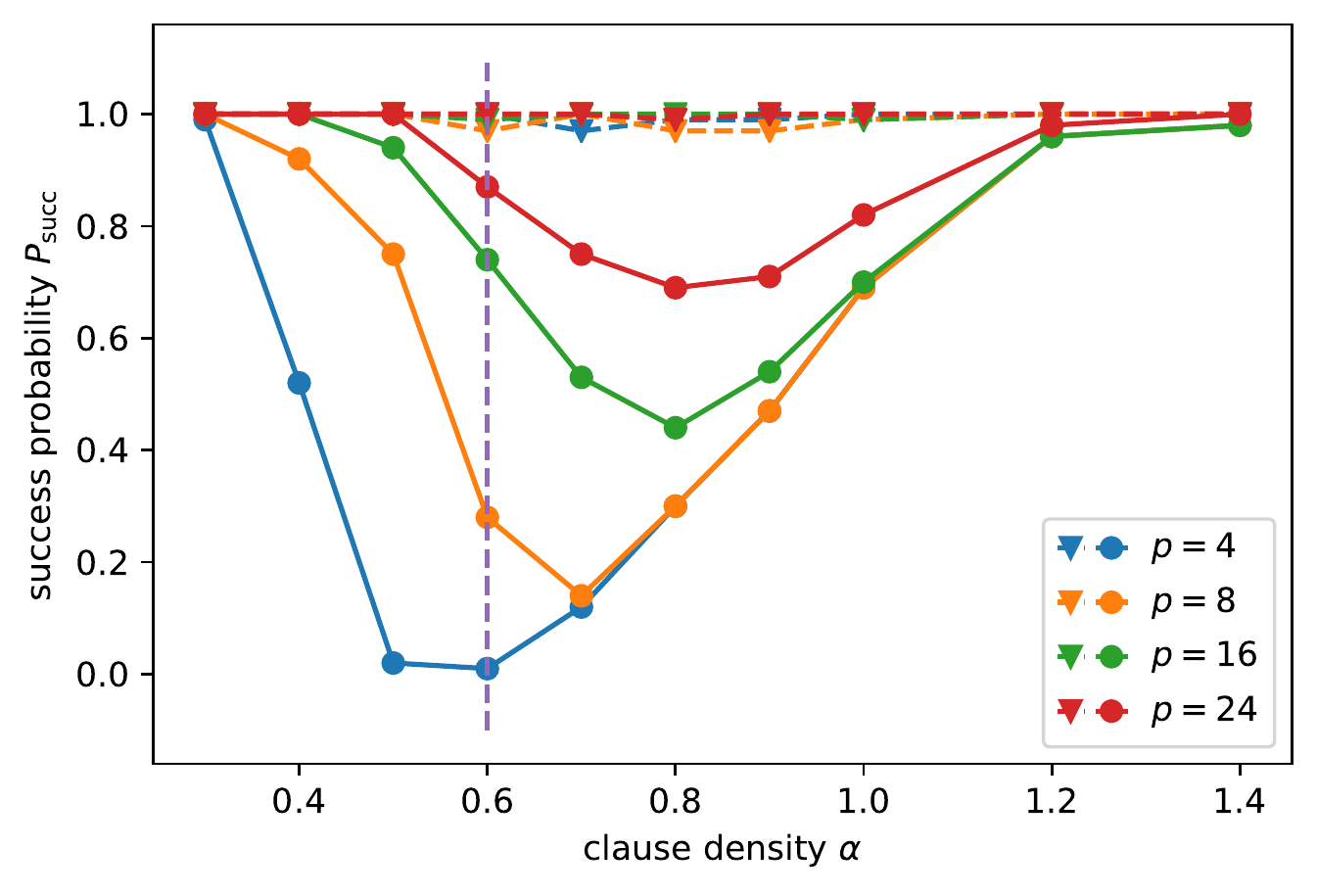}
		\caption{Comparison of the success probability between the QAOA (solid lines) and the ab-QAOA (dashed lines) for solving $10$-variable $1$-$3$-$\mathrm{SAT}^+$ problems as a function of the clause density $\alpha$, where the increment of $\alpha$ is $0.1$ from $\alpha=0.3$ to $1$ and is $0.2$ from $\alpha=1$ to $1.4$. Each point is an average over $100$ random instances and the system size is $n = 10$. The vertical lines represent the critical clause density $\alpha_\mathrm{c}=0.6$. The QAOA cannot solve the NP-complete $1$-$3$-$\mathrm{SAT}^+$ problems even at level $24$.  The ab-QAOA, obtains virtually all the right answers at level $4$. }\label{fig:succ_probability}
	\end{figure}

  For the QAOA there is indeed an easy-hard-easy pattern in the quantum cost, as seen in Fig.~\ref{fig:succ_probability}, where $10$ variables are considered. The problems near the SAT-UNSAT transition point are the hardest to solve. In sharp contrast, the ab-QAOA can solve these problems with near-perfect success probability in only level $4$. The easy-hard-easy pattern of the quantum cost is less evident in ab-QAOA.

	\subsection{Residual energy and infidelity}
	\label{subsec:residual}

\begin{figure}[htb]
		\centering
		
        \includegraphics[scale=0.43]{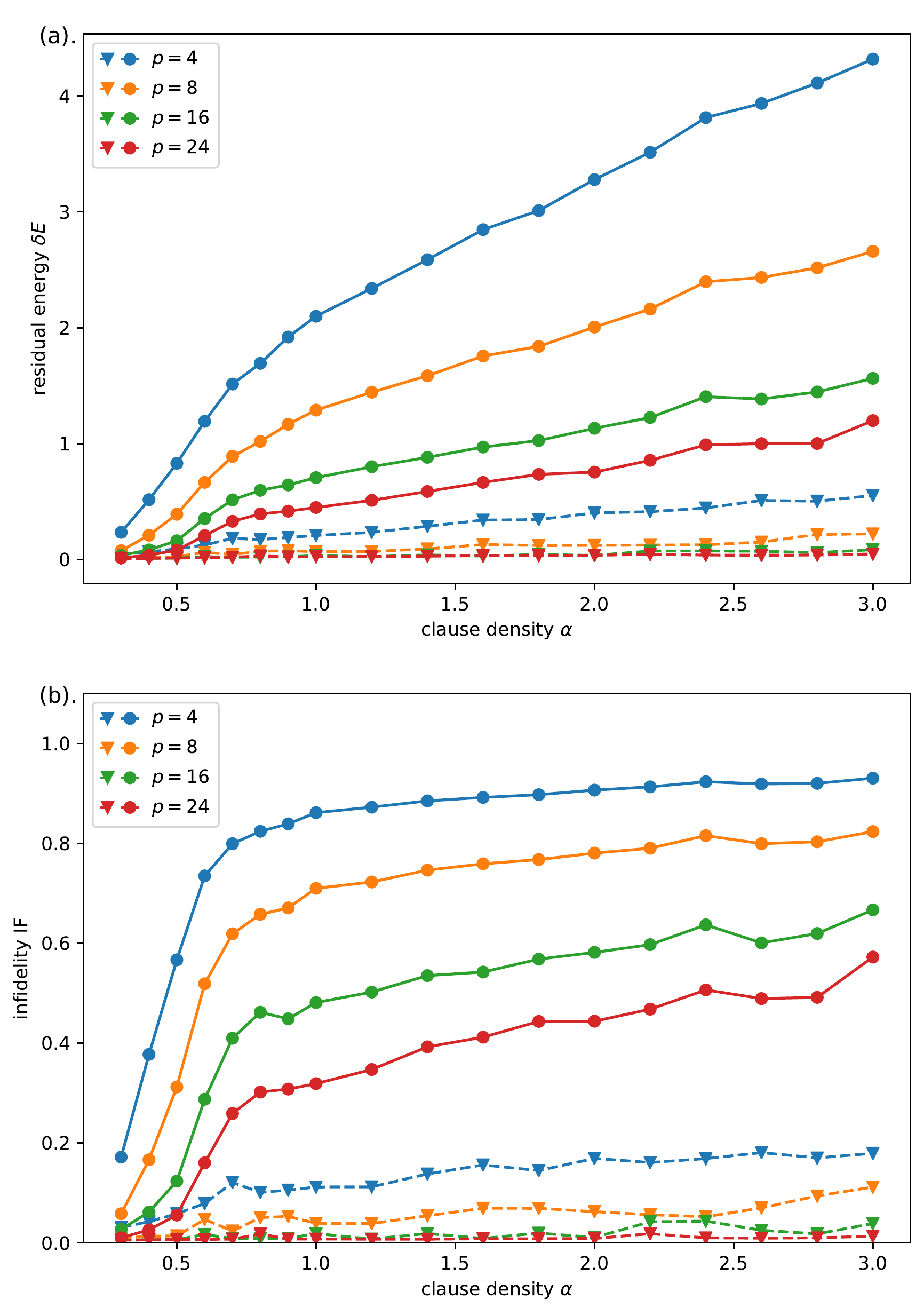}
		\caption{Comparison of the residual energy (a) and infidelity (b)  of QAOA (solid lines) to that of the ab-QAOA (dashed lines) for the $10$-variable Max-1-3-SAT$^+$ problem. Results are given for levels $p=4,8,16,24$ as a function of the clause density $\alpha$. Each point is an average over $100$ random instances and n = 10. The Modified Trotterized quantum annealing method is applied to the QAOA and ab-QAOA. The classical optimizer is Adam~\cite{adam}, which has been used in conjunction with a real device~\cite{Rydberg_qaoa3}. The performance of the ab-QAOA in level $4$ is superior to that of the QAOA in level $24$. The reachability deficits are much less evident in level $8$ for the ab-QAOA than for the QAOA in level $24$.} \label{fig:energy_fidelity}
	\end{figure}

	For the Max-$1$-$3$-$\mathrm{SAT}^+$ problems, we use the residual energy $\delta E$ and the infidelity $\mathrm{IF}$ as functions of the clause density $\alpha$ for the QAOA and the ab-QAOA for benchmarking purposes. The residual energy is defined as, 
    \begin{align}
		\delta E=\langle H_\mathrm{C} \rangle-E_\mathrm{g},
    \end{align}
	where $\langle H_\mathrm{C} \rangle$ is the expectation value of the cost Hamiltonian output from the QAOA or the ab-QAOA and $E_\mathrm{g}$ is the ground energy of $H_\mathrm{C}$. The infidelity is defined as,
 
    \begin{align}
     \mathrm{IF}=1-\sum_l|\langle\psi_\mathrm{f}|\psi_\mathrm{g}^l\rangle|^2,
    \end{align}
    where $|\psi_\mathrm{f}\rangle$ is the output state of the QAOA ($|\psi_\mathrm{f}^\mathrm{s}\rangle$) or the ab-QAOA($|\psi_\mathrm{f}^\mathrm{ab}\rangle$) and 
	$|\psi_\mathrm{g}^l\rangle$ is the product ground state of $H_\mathrm{C}$ with $l$ labeling the degeneracy.  The results of $10$-variable problem instances are shown in Fig.~\ref{fig:energy_fidelity}. 
	
	For both algorithms, increasing the level $p$ reduces both error measures.  However, the QAOA can not solve the high clause density problems effectively even at level $24$, which is of course far beyond the capabilities of current NISQ devices~\cite{sycamore_qaoa}.  
	As for the ab-QAOA, even a level-$4$ ab-QAOA is better than the QAOA with level $24$, which implies that the reachability deficits~\cite{reachabiilty_deficits_sat} pose fewer problems for the ab-QAOA than for the QAOA. In level $8$, an accurate state can be found by ab-QAOA with small residual energy and infidelity. For different system sizes $n$, the residual energy and infidelity are also presented in Appendix~\ref{sec:ef_4}. The maximal clause density of the $10$-variable problems considered in the present work is $11$, and the numerical results beyond $\alpha=3$ are presented in Appendix~\ref{sec:ef_4}, where the reachability deficits are still much less evident in the ab-QAOA. 
 
    The main point here concerns the easy-hard transition. It is remarkable how poorly the QAOA does at high $\alpha$.  At $p=4$ and $\alpha=3$, the infidelity is approaching unity, its maximum value. At $p=24$, $\mathrm{IF}$ still exceeds $0.5$. In contrast, the ab-QAOA result at $p=4$ is less than $0.2$ and at $p=24$ it is less than $0.02$.
 
	\subsection{Quantum cost}\label{subsec:qc}
\begin{figure}[htb]
		\centering
		\includegraphics[scale=0.43]{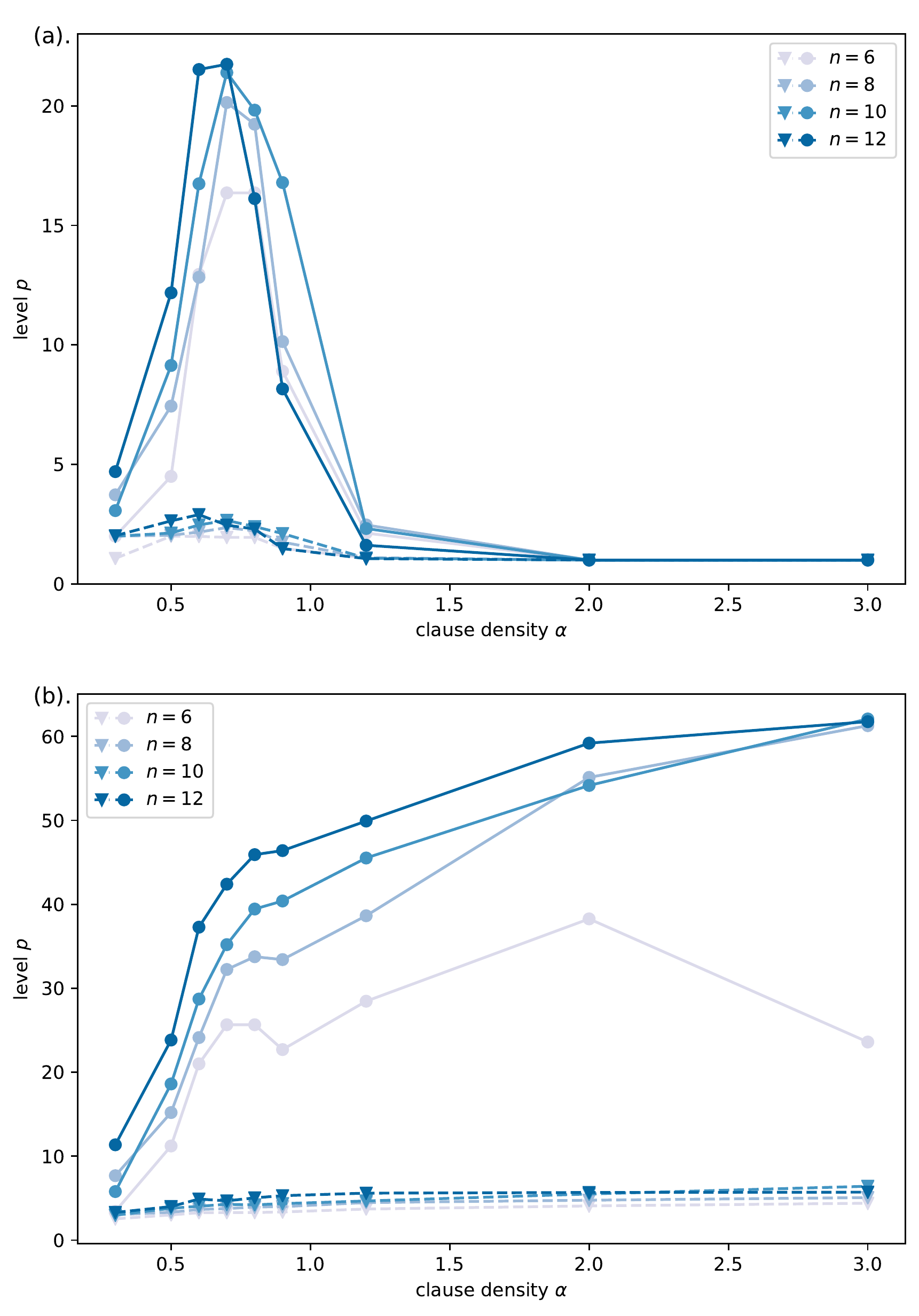}
		\caption{Levels needed to solve $1$-$3$-$\mathrm{SAT}^+$ problems (a) and Max-$1$-$3$-$\mathrm{SAT}^+$ problems (b) for QAOA(solid lines) and ab-QAOA(dashed lines), by the criterion $\mathrm{IF} \leq 0.1$. If the problem is not solved at level 64, then 64 is recorded. Hence the point in the upper right corner of (b) means that the QAOA did not solve the case $\alpha = 3.0$. Different colors represent the different system sizes from $n=6$ to $n=12$. Each point is the average over $100$ problem instances except for $n=12$ where only $50$ problem instances are calculated. The clause densities here cover the hard region problems described in the main text.}\label{fig:cost}
	\end{figure}

	In Fig.~\ref{fig:cost}, we plot the level $p$ required to solve, exactly or approximately, the Max-$1$-$3$-$\mathrm{SAT}^+$ problems and $1$-$3$-$\mathrm{SAT}^+$ problems as functions of $\alpha$. For the Max-$1$-$3$-$\mathrm{SAT}^+$ problem instance, we calculate the infidelity from level $1$ to level $8$ with an increment of $1$ and from level $8$ to level $64$ with an increment of $8$ for the QAOA and record the level at which the inequality $\mathrm{IF}\leq0.1$ is first satisfied, so that an approximate solution has been achieved. Since $p=64$ is the highest level we calculate, in those instances that $\mathrm{IF}$ is still larger than $0.1$ in level $64$, we record the final level as $64$. For the $1$-$3$-$\mathrm{SAT}^+$ problems, record the level where the QAOA is successful.   
 
    For the ab-QAOA, levels smaller than $8$ are enough to solve the problems above and an exact $p$ can be easily obtained. For the QAOA, when $\alpha$ is small ($\alpha = 0.3$), $8$ levels are enough to give the solutions. However, when $\alpha$ is large, due to the existence of reachability deficits~\cite{reachabiilty_deficits_sat} for some Max-$1$-$3$-$\mathrm{SAT}^+$ problems, the needed $p$ value increases rapidly and at $\alpha>2.0$ the value is so large ($p>64$) that we were not able to determine it. For simplicity, we approximate it as $64$.

    In Fig.~\ref{fig:cost}, it is clear that the quantum cost of the QAOA exhibits an easy-hard pattern for the Max-$1$-$3$-$\mathrm{SAT}^+$ problems and an easy-hard-easy pattern for the $1$-$3$-$\mathrm{SAT}^+$ problems, while these patterns are far less evident in the ab-QAOA, where again the levels are taken from $1$ to $8$ with an increment of $1$ and from $8$ to $24$ with an increment of $8$. As seen in Fig.~\ref{fig:cost}, the problem hardness increases with the system size, thus more levels are needed to achieve a given accuracy, which is also consistent with Fig.~\ref{fig:energy_fidelity4} in Appendix~\ref{sec:ef_4}, where both the residual energy and infidelity increase with $n$. The convergence at level $64$ in Fig.~\ref{fig:cost}(b) for different $n$ implies level $64$ is not enough to solve the problems for the QAOA. 
    
    For the QAOA results of $n=6$ and $\alpha=3$ in Fig.~\ref{fig:cost}(b), the levels needed for solving the problems here are smaller compared with other $\alpha$ in $n=6$, which means the problems are easier. This is because the maximal clause density for $n=6$ is $10/3$,  when $\alpha$ achieves its maximal value, the cost Hamiltonian tends to the identity matrix, for which it is easy to find the ground state. We also checked the gap between the ground state and the first excited state and found that the gap is amplified when $\alpha$ is close to the maximal value. This leads to the improved performance of the QAOA~\cite{ab_qaoa}. This analysis can also be applied to the results of $n=10$ and $\alpha=11$ in Fig.~\ref{fig:energy_fidelity4}.
    
    The level $p$ is related to the quantum cost as follows. 
 The number of quantum gates needed to achieve a given accuracy in QAOA or ab-QAOA is proportional to $p^2$, as shown in detail in Ref.~\cite{ab_qaoa}. This dependence follows from the observations that at level $p$ there are $2p+1$ gradients to be computed for each component of the QAOA variational parameters, and $2p$ operators to be applied for each gradient calculation. Thus we may write the number of gates as $O(N_\mathrm{con}p^2)$, where $N_\mathrm{con}$ is the number of iterations needed for a specified accuracy. In the numerical simulations of $n=10$, we found that $N_\mathrm{con}$ is around $30$ in the QAOA and about $25$ in the ab-QAOA. For the hard-region $1$-$3$-$\mathrm{SAT}^+$ problems, $\alpha = 0.6,0.7,0.8$, the levels needed in QAOA are $17,22,20$ compared with $3$ in the ab-QAOA.  By using $N_\mathrm{con}p^2$ as the measure of computation time, we conclude that a $50$-fold speedup is achieved for ab-QAOA over QAOA. As for the hard-region Max-$1$-$3$-$\mathrm{SAT}^+$ problems, $\alpha=0.9,1.2,2,3$, the QAOA levels are about $41,46,55,63$ while the ab-QAOA levels are $5,5,6,7$, so the ab-QAOA can achieve roughly a $90$-fold speedup on average. 
    
    \subsection{$R$ dependence}
    
\begin{figure}[htb]
		\centering
		\includegraphics[scale=0.43]{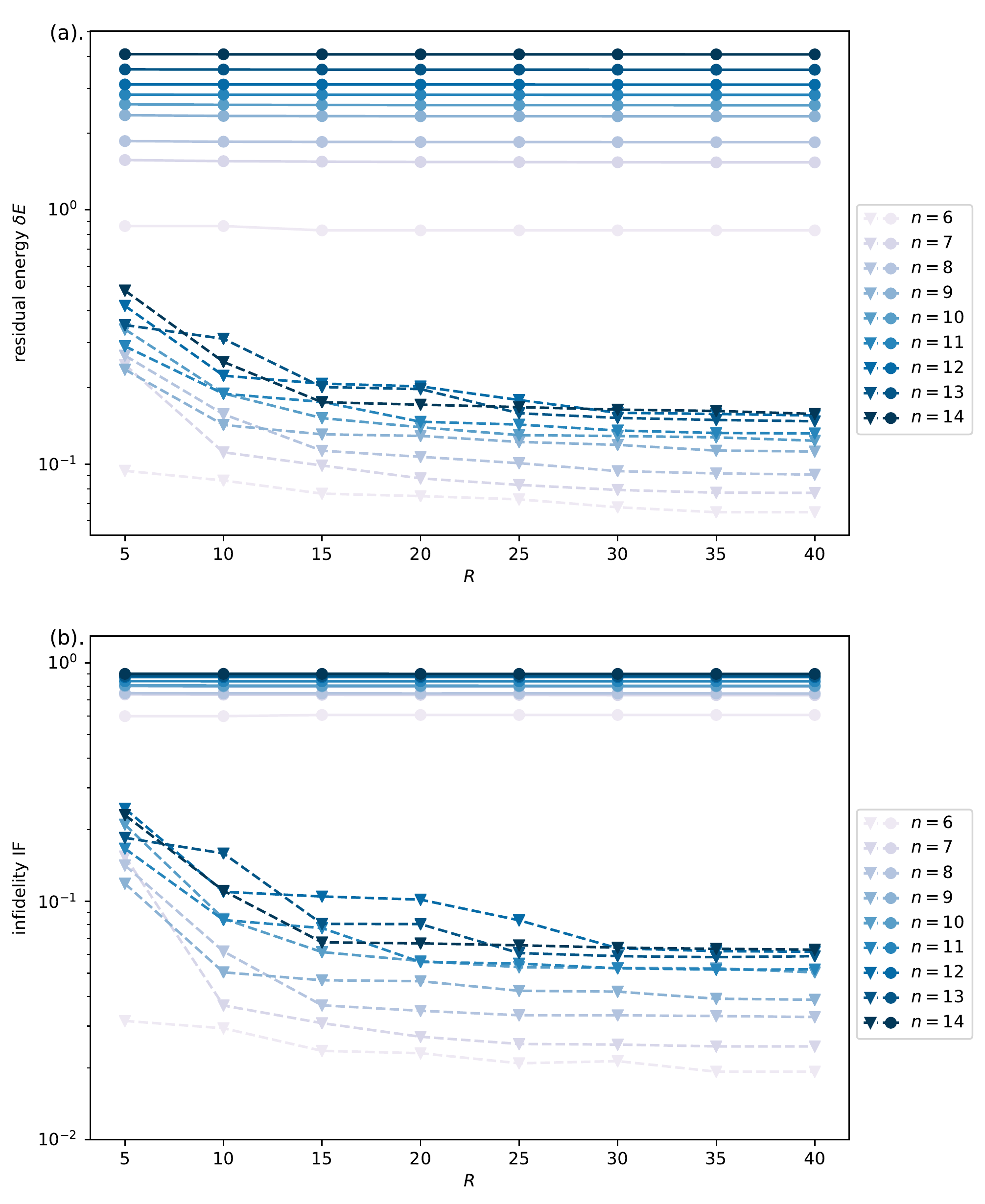}
		\caption{Comparison of the residual energy (a) and the corresponding infidelity (b) between the QAOA (solid lines) and the ab-QAOA (dashed lines) as a function of the sample parameter $R$. 
		The optimization begins from the modified TQA method with level $p=8$. Both the residual energy and infidelity are relatively insensitive to $R$ in the QAOA.  Each point is an average over $50$ random Max-$1$-$3$-$\mathrm{SAT}^+$ problem instances with clause density $\alpha=3$.}\label{fig:R_depend}
	\end{figure}

\begin{figure}[htb]
		\centering
		\includegraphics[scale=0.55]{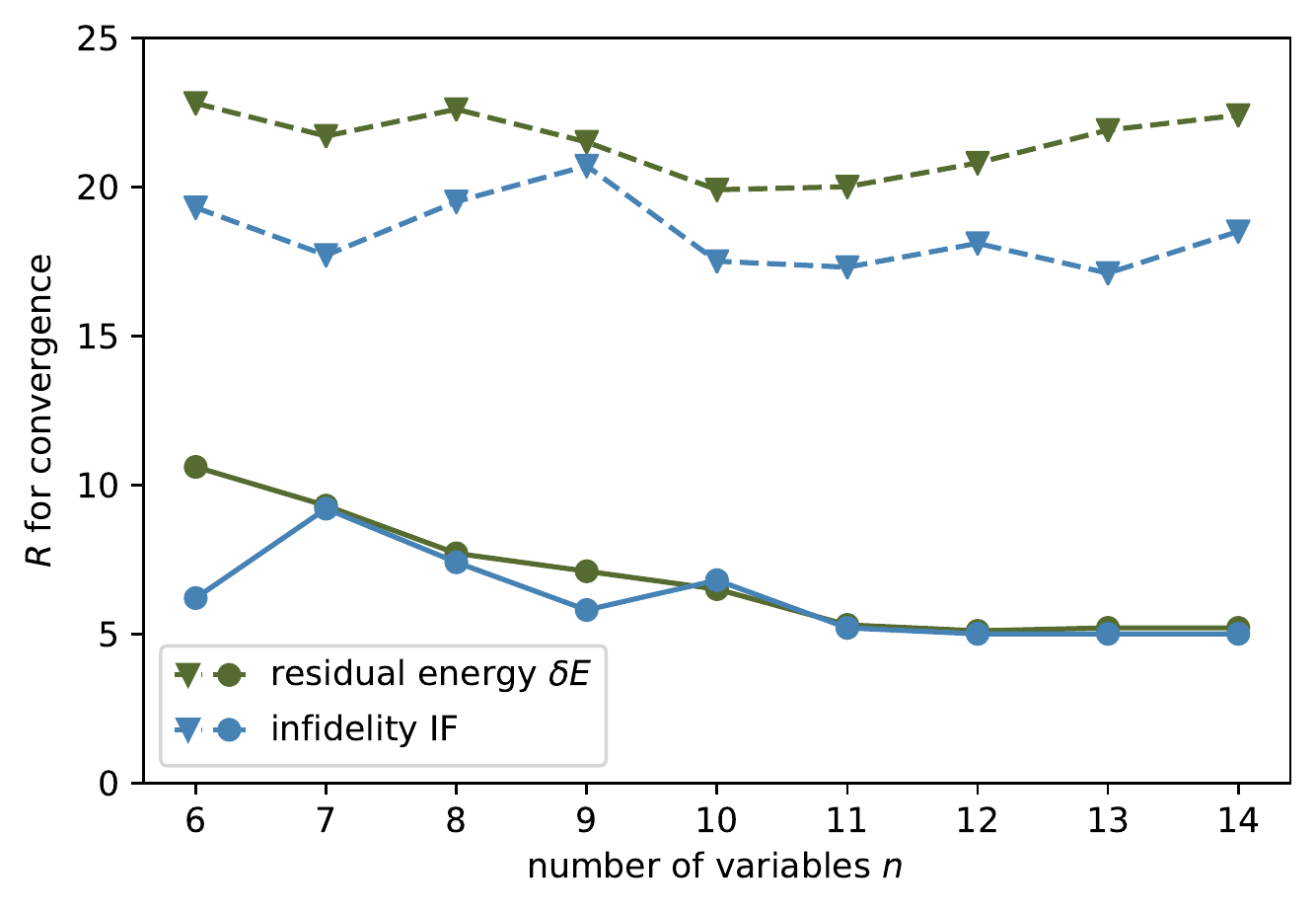}
		\caption{Comparison of the convergence of the residual energy (green curves) and the corresponding infidelity (blue curves) in the QAOA (solid lines) and the ab-QAOA (dashed lines) as a function of $n$, the number of variables. The ab-QAOA needs $20\sim25$ samples while the QAOA needs $5\sim10$ samples. The problem instances are the same as those in Fig.~\ref{fig:R_depend}.}\label{fig:R_depend_con}
	\end{figure}

    In the implementation of the ab-QAOA, the optimization starts from $R$ initial points in parallel, as defined in Algorithm~\ref{alg:tqa} and Fig.~\ref{fig:abqaoa}. Following the optimization the point with the lowest energy is selected out. A larger $R$ means more points are covered in the energy landscape and this leads to a smaller residual energy. It will ultimately be necessary to understand exactly how large $R$ should be to demonstrate the advantages of the ab-QAOA.  It is also important that $R$ scales favorably with the problem size $n$, since there is a danger if  the quantity $R$ grows fast with $n$. 
    To investigate this issue, we have computed how the residual energy $\delta E$ and the corresponding infidelity $\mathrm{IF}$ vary with different $R$. The details of $\delta E$ and $\mathrm{IF}$ as a function of $R$ are shown in Fig.~\ref{fig:R_depend}. If the relative changes in the lowest residual energy or the lowest infidelity are less than $10^{-2}$ when increasing $R$, the current $R$ is recorded as the convergent value, which is shown in Fig.~\ref{fig:R_depend_con}. One sees that $\delta E$ and $\mathrm{IF}$ in the QAOA converge quickly starting from $R=5$.

    It is observed that even with $R=5$, the ab-QAOA still outperforms the QAOA in Fig.~\ref{fig:R_depend} .
    It is observed in Fig.~\ref{fig:R_depend_con} for all the system sizes considered here, the values of $R$ from which the convergence starts do not change significantly with $n$. 
    These two observations imply that a very accurate guess of the ground state is not necessary when encoding the bias fields at the initial stage of the optimization. Based on this, the convergent values of $R$ should not change dramatically with the system size in other cases.
 
	\section{Comparative Analysis of QAOA and ab-QAOA}
	\label{sec:analysis}
	
	\subsection{Introduction}
	\label{subsec:introduction}
	The QAOA is a fairly general algorithm in that it derives from the QAA~\cite{qaoa_farhi,qaoa_lukin}. The motivation and performance guarantees for the latter come from the adiabatic theorem, which applies to all final states.  Hence the QAOA can overall be expected to apply to any problem Hamiltonian, with the usual caveats about small gaps~\cite{qaoa_lukin}. This raises the possibility that if the problem Hamiltonian is known to have a special structure, one could modify the QAOA to take advantage.  This is the case for the Ising-model problem Hamiltonians.  The ground states are very special: the computational basis states. The ab-QAOA is an algorithm that is targeted towards just these states, and this explains its superiority to the QAOA for this class of problems. As already noted, the limitation to Ising model Hamiltonians is surprisingly unrestrictive, at least in the context of classical combinatorial optimization.
	
	To understand the targeted character of the ab-QAOA, consider the typical initial state $|\psi_0^\mathrm{s}\rangle = |-\rangle^{\otimes n}$ and the general Ising-model ground state $|\psi_\mathrm{g}\rangle = |s_1\rangle \otimes |s_2\rangle \otimes |s_{n}\rangle $.  Here $s_j = 0 $ or $1$ and $\{s_j\}_{j=1}^n$ encodes the solution to the optimization problem.  The ab-QAOA wavefunction is reinitialized to,
	\begin{align}
        \begin{split}
        	    &|\psi_{0}^{\mathrm{ab}}(\vec{h})\rangle=\tilde{R}_y(\vec{d}\,) |-\rangle^{\otimes n},\\
	       &=\bigotimes_{j=1}\left[\frac{(\cos\frac{d_j}{2}+\sin\frac{d_j}{2})|0\rangle_j-(\cos\frac{d_j}{2}-\sin\frac{d_j}{2})|1\rangle_j}{\sqrt{2}}\right].
        \end{split}\nonumber
	\end{align}
	When a quite accurate approximation of $|s_j\rangle$ with $h_j$ is obtained (this means $h_j>0$ or equivalently $d_j>0$ if $s_j=0$ and $h_j<0$ or $d_j<0$ if $s_j=1$), then the overlap between $|s_j\rangle$ and the ab-QAOA initialized state on qubit $j$ is $\sqrt{(1+|\sin d_j|)/2}$, which means the probability amplitude of the solution state is amplified in the reinitialization procedure. 
     This is in stark contrast to the QAOA, where the initial state is fixed. This procedure is only practical  because of the absence of entanglement in the solution state.  The evolution operator itself evolves, not only the wavefunction and the operator schedule.
    
    The question is then whether this leads only to a small incremental improvement in computational power or whether there is a deeper advantage for the ab-QAOA.  One may conjecture from the above discussion that the reinitialization will have the  consequence that the ab-QAOA wavefunction remains in a part of the Hilbert space with relatively low entanglement compared to the QAOA.  For Ising problems this can be an advantage, since both the initial and final states are product states.  
    
    Furthermore, we know that if the wavefunction explores all of Hilbert space in a uniform fashion, then the barren plateau phenomenon will reduce the effectiveness of the algorithm~\cite{barren_plateaus}.  From the results on the SAT problems in the Sec.~\ref{sec:result} we know that the gap between the QAOA and the ab-QAOA is greatest when the barren plateaus are most evident.
    Similarly, there is numerical evidence that the scaling behavior with system size of the accuracy is better in the ab-QAOA than in the QAOA~\cite{ab_qaoa}, it will be interesting to investigate the barren plateau phenomenon with ab-QAOA. We leave this for the future work.
    The remainder of this section is devoted to investigating the conjecture that there is reduced entanglement in the ab-QAOA and this is responsible for 
    \begin{figure*}[htb]
		\centering	
		\includegraphics[scale=0.43]{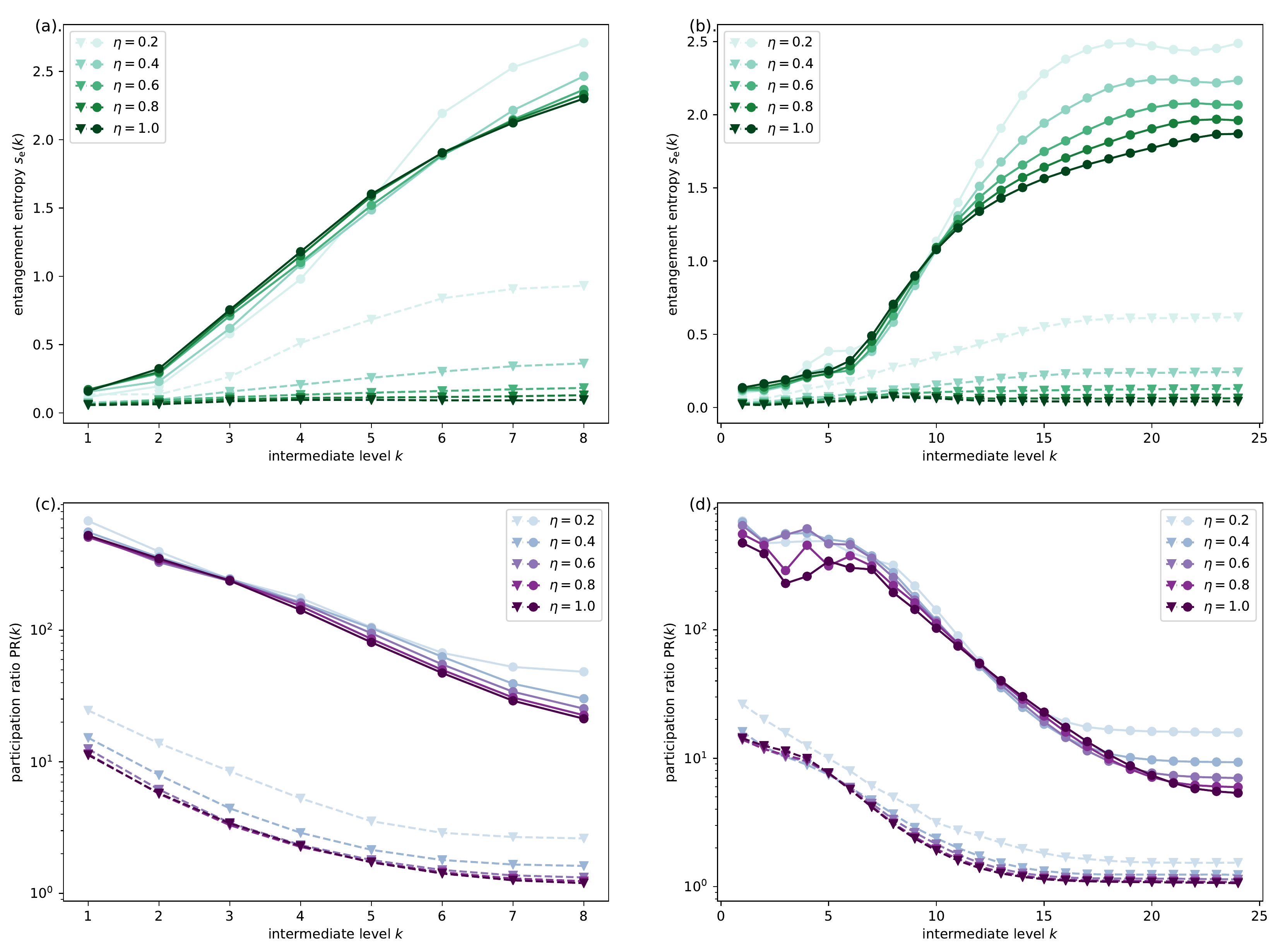}
\caption{The dynamical evolution of the entanglement entropy $s_\mathrm{e}(k)$ (a, b) and the participation ratio (c, d) in the $10$-variable Max-1-3-SAT$^+$ problem for the QAOA (solid lines) and the ab-QAOA (dashed lines) at levels $p=8$ (a, c) and $p=24$ (b, d). The parameter $k$ is the ''time'' elapsed in the quantum evolution, as defined through Eq.~\eqref{eq:intermediate_state}. 
The parameter $\eta$ is the dynamical parameter in the optimization, as defined in Eq.~\eqref{eq:defn.eta}. 
The problem instances are the same as those with $\alpha=3$ in Fig.~\ref{fig:energy_fidelity}. The ab-QAOA is less entangled and more localized because of the bias fields in the mixing Hamiltonian.}\label{fig:entangle_pr}
	\end{figure*}
    its power. We compute the entanglement entropy, the participation ratio, and the annealing entropy.
    
\subsection{Entanglement entropy and participation ratio}\label{sec:result_entangle}
	We first establish the fact that the entanglement entropy is much lower in the ab-QAOA evolution than in that of the QAOA.  Since entanglement is independent of the basis chosen and we contend that the association with the computational basis is paramount, we also compute the participation ratio to show that the reduced entanglement comes from being close to product states in this preferred basis. The latter is a form of many-body localization~\cite{qaa_catalyst}.
	
	We need to see how these quantities evolve as the system moves from the initial state to the approximate ground state.  The natural "time" parameter in the evolution is the number of times the unitary operators have been applied. More precisely, when we prepare the final state $|\psi_\mathrm{f}\rangle$, either $|\psi_\mathrm{f}^\mathrm{s}\rangle$ in Eq.~\eqref{eq:qaoa} or $|\psi_\mathrm{f}^\mathrm{ab}\rangle$ in Eq.~\eqref{eq:ab_qaoa}, the system undergoes a $p$-step discrete time evolution driven by the unitary operators,
	\begin{align}
		U_k=\mathrm{e}^{-i\beta_{k} H_\mathrm{M}} \mathrm{e}^{-i\gamma_kH_\mathrm{C}},
	\end{align}
	which are applied sequentially.  Here $U_k$ can represent both QAOA $U_k^\mathrm{s}$ and ab-QAOA $U_k^\mathrm{ab}$ by setting $H_\mathrm{M}=H_\mathrm{M}^\mathrm{s}$ or $H_\mathrm{M}=H_\mathrm{M}^\mathrm{ab}$. When the superscripts are dropped, the corresponding operators and states represent both QAOA and ab-QAOA. The intermediate states $|\psi_{k}\rangle$ after $k$ steps satisfy,
	\begin{align}
		|\psi_k\rangle=U_k|\psi_{k-1}\rangle, \label{eq:intermediate_state}
	\end{align}
	where $|\psi_{0}\rangle$ is the initial state of the QAOA or ab-QAOA. We will use $k$ to observe the course of the optimization process.
	
	We can gain further information about the evolution by investigating how quickly the optimization converges. For this purpose, we define a parameter $\eta$ as follows.  Let $N_\mathrm{con}$ be the number of iterations needed for the convergence for a given initial $(\vec{\gamma},\vec{\beta},\vec{h})$.  Quantities such as the entanglement entropy and the participation ratio at the $\mathcal{N}^\mathrm{th}$ step of the iteration can then be indexed using a variable $\eta$, 
	\begin{eqnarray}
	\eta = \frac{\mathcal{N}}{N_\mathrm{con}},     \quad \quad 0 \leq \eta \leq 1 \; .
	\label{eq:defn.eta}
	\end{eqnarray}
	%

	To compute the entanglement entropy, the total system $\rho_k=|\psi_k\rangle\langle\psi_k|$ is divided into $2$ parts, $\mathrm{A}$ and $\overline{\mathrm{A}}$, and the entanglement entropy $s_\mathrm{e}(\rho_{k,\mathrm{A}})$ associated with this bipartition is, 
	\begin{align}
		s_\mathrm{e}(\rho_{k,\mathrm{A}})=-\mathrm{Tr}(\rho_{k,\mathrm{A}} \log_2 \rho_{k,\mathrm{A}}),
	\end{align}
	where $\rho_{k,\mathrm{A}}$ is the reduced density matrix for $\rho_k$ in part A. We consider all the possible bipartitions, take an average over them and record the average entanglement entropy as $s_\mathrm{e}(k)$. For the $10$-qubit system in this paper, the average entropy $s_\mathrm{e}(k)$ satisfies,
	\begin{align}
		0 \leq s_\mathrm{e}(k)\leq 3.7769,
	\end{align}
	where the lower bound (upper bound) is reached when subsystem $\mathrm{A}$ is a product state (fully mixed state) in all possible bipartitions.

	In Figs.~\ref{fig:entangle_pr} (a) and (b), the entanglement entropy of the intermediate states $s_\mathrm{e}(k)$ is plotted. The dynamical behavior of $s_\mathrm{e}(k)$ in the optimization with $\eta\in\{0.2,0.4,0.6,0.8,1\}$ is also shown. The results in Figs.~\ref{fig:entangle_pr} (a) and (b) are for the problem instances with $\alpha=3$, the same as those in Figs.~\ref{fig:energy_fidelity}. Numerical results for $\alpha=0.8$ can be found in Appendix~\ref{sec:res_a8}. The results clearly  show that the bias fields $\vec{h}$ cause the states in the evolution of the ab-QAOA to be much less entangled than those of the QAOA. This enhances the speedup in finding the ground state, since the ground state can be represented by localized product states. For the same reason, the optimization process reduces the entanglement further. As for the QAOA, it is interesting that there seems to be a critical $k \approx 5$ when $p=8$ and $k \approx 10$ when $p=24$ only beyond which can the optimization reduce the entanglement. The critical $k$ is close to that in Sec.~\ref{sec:result_annealing}. In the big $p$ limit, as $k$ increases, the entanglement entropy $s_\mathrm{e}(k)$ must eventually decrease as the optimization brings the QAOA output state closer to the product ground state.

	Entanglement itself is basis-independent, so we need to probe a little deeper to find the reason for the difference in entanglement between the QAOA and ab-QAOA.  For this we compute the participation ratio
	defined as~\cite{qaa_catalyst},
	\begin{align}
		\mathrm{PR}(k)=\left[\sum_\phi|\langle\phi|\psi_{k}\rangle|^4\right]^{-1},
	\end{align}
	where $\phi$ runs over the computational basis. This is a (necessarily basis-dependent) measure of many-body localization.  If $\mathrm{PR}(k)=1$, $|\psi_{k}\rangle$ is completely localized in this basis, \textit{i.e.,} it is a computational basis state.  It is maximally delocalized when $\mathrm{PR}(k)=2^{10}>>1$ in our 10-qubit system. As shown in Figs.~\ref{fig:entangle_pr} (c) and (d), combined with Figs.~\ref{fig:entangle_pr} (a) and (b), the ab-QAOA is more localized.
	
	We conclude from these results that the states involved in the ab-QAOA evolution cleave closely to the computational basis states and are thus more localized.  In this part of phase space the energy optimization guides the state quickly to the optimal one.  In contrast, the QAOA evolution appears to wander into a more delocalized region and thus converges far more slowly.

	\subsection{Annealing entropy}\label{sec:result_annealing}
	We denote the $k$-dependent orthonormal basis in which the unitary time-evolution operator $U_k$ is diagonal by $\{|\theta_k\rangle\}$:
	\begin{align}
		U_k|\theta_k\rangle=\mathrm{e}^{-i\theta_k}|\theta_k\rangle.\label{eq:intermediate_basis}
	\end{align}
	In an ideal adiabatic process a system that starts in the state $|\theta_k\rangle$ also finishes there. Following Ref.~\cite{annealing_entropy}, we define the annealing entropy $s_\mathrm{a}(k)$ to describe the deviation from perfect adiabaticity:
	\begin{equation}
	\begin{split}
		s_{\mathrm{a}}(k) & = -\sum_{\theta_k}	F_k(\theta_k)\log_2 F_k(\theta_k), \\
		F_k(\theta_k)
		& =  |\langle\theta_k|\psi_k)\rangle|^2,
    \end{split}
	\end{equation}
	where $F_k(\theta_k)$ is non-negative. For ideal adiabaticity $F_k(\theta_k)$ has only two values, $0$ and $1$, so the annealing entropy is $0$. The less the annealing entropy, the more adiabatic the discrete evolution. 

    \begin{figure}[htb]
		\centering
		\includegraphics[scale=0.43]{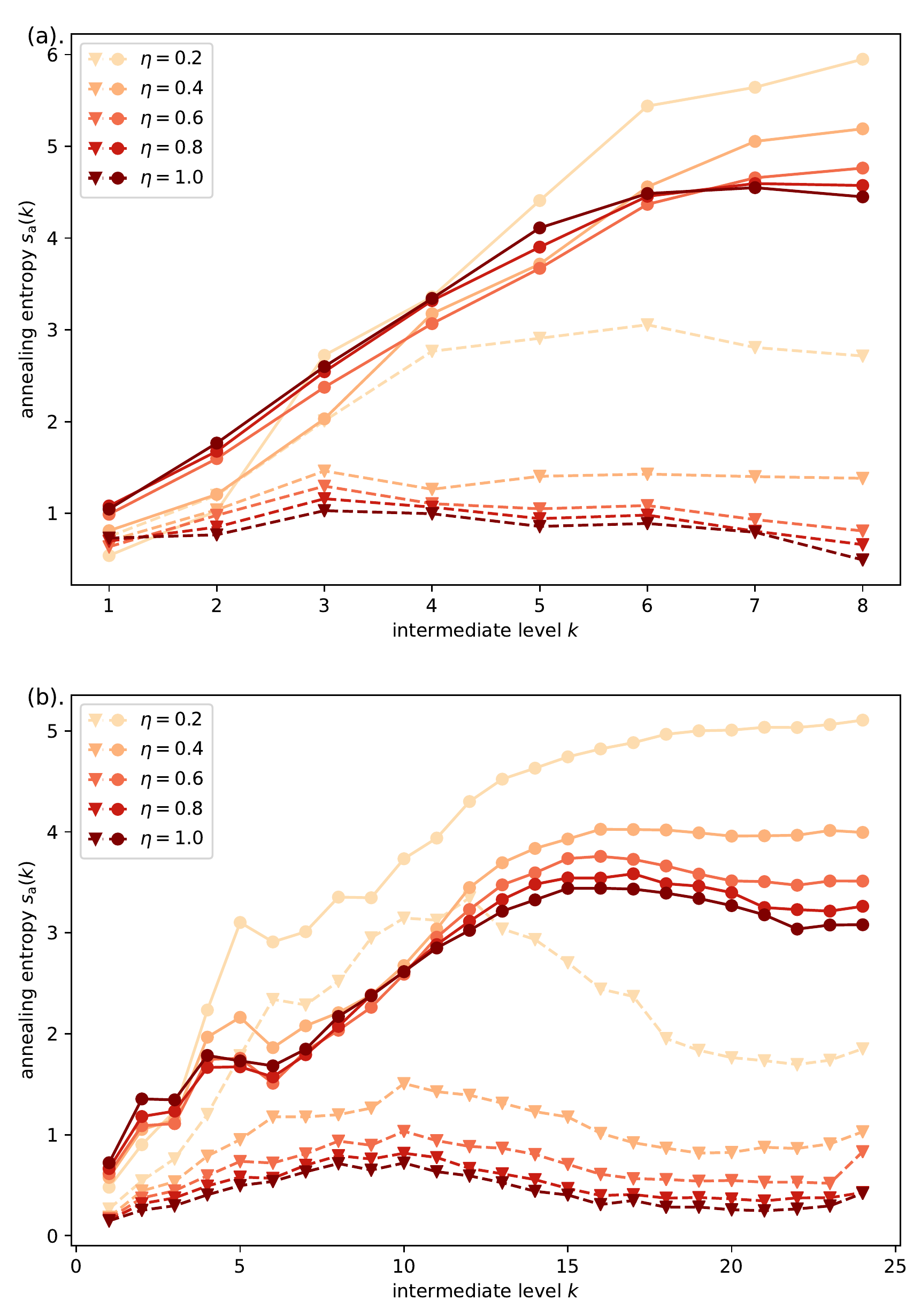}
		
		\caption{The dynamical evolution of the annealing entropy $s_{\mathrm{a}}(k)$ of the QAOA (solid lines)  and the ab-QAOA (dashed lines) at level $p=8$ (a) and level $p=24$ (b) for the $10$-variable Max-1-3-SAT$^+$ problem. The problem instances are the same as those in Fig.~\ref{fig:energy_fidelity} that have $\alpha=3$. The much smaller annealing entropy of the ab-QAOA shows that it is more adiabatic than the QAOA.}
		\label{fig:annealing}
	\end{figure}
  
   We calculate the intermediate $s_{\mathrm{a}}(k)$ for the hard-region Max-$1$-$3$-$\mathrm{SAT}^+$ problems with $\alpha=3$ and show the results in Fig.~\ref{fig:annealing} for $p = 8$ (a) and $ p = 24$ (b). 
   The dynamical parameter $\eta$ runs over the set $\{0.2,0.4,0.6,0.8,1\}$. It is evident that the evolution of the ab-QAOA is much closer to the adiabatic limit than the evolution of the QAOA. The modification of the starting state in the ab-QAOA promotes this reduction. For the QAOA, there is a critical intermediate level, only beyond which can the optimization reduce the adiabaticity. The results of calculations for the hard-region $1$-$3$-$\mathrm{SAT}^+$ problems with $\alpha=0.8$ are given in Fig.~\ref{fig:annealing2} in Appendix~\ref{sec:res_a8} and show similar patterns.

	\section{Optimization-free ab-QAOA}\label{sec:of_abqaoa}
	As explained above and in Ref.~\cite{ab_qaoa}, in the ab-QAOA the true ground state can actually be generated by $H_\mathrm{M}^\mathrm{ab}$ itself if the variational parameters $\vec{\gamma},\vec{\beta}$  and the bias field parameters $\vec{h}$ are correctly chosen.  This suggests the idea of only using $\vec{h}$ to construct a product "bias state", using the fact that $h_j=+1$ leads to the state $|0\rangle_j$ while $h_j=-1$ leads to the state $|1\rangle_j$.  This leads to a kind of "greedy" version of the ab-QAOA that is free of classical optimization.  This "optimization-free QAOA" is shown in detail in Algorithm~\ref{alg:of_abqaoa}.	In our calculation, $R=10$ and $\delta t=0.6$ with $\delta t$ defined in Algorithm~\ref{alg:of_abqaoa}. The source codes are available in~\cite{codes}. 
	\begin{algorithm}
		\caption{Optimization-free ab-QAOA}
		\label{alg:of_abqaoa}
		\SetKwInOut{Return}{Return}
		\KwIn{target level $p$, $R$ randomly-generated bias fields $\{\vec{h}^{1r}\}$ with components to be $1$ or $-1$.}
		\KwOut{final energy $E_p$ in level $p$ and the corresponding state.}
		\For{$p^\prime=1$ \KwTo $p$}{
			Initialize $\vec{\gamma}$ and $\vec{\beta}$ according to
			\begin{align}
				\gamma_{k}&=\frac{k-1}{p^\prime} \delta t,\nonumber\\
				\beta_{k}&=(1-\frac{k-1}{p^\prime})\delta t.\nonumber
			\end{align}\\
			\For{$r=1$ \KwTo $R$}{
				1. Prepare the state  $|\psi_\mathrm{f}^\mathrm{ab}\rangle$ with $(\vec{\gamma},\vec{\beta},\vec{h}^{p^\prime r})$, measure $\langle Z_j \rangle$ for each qubit.\\
				2. Update $\vec{h}^{p^\prime r}$ according to 
				\begin{align}
					h_j^{p^\prime r} = h_j^{p^\prime r}-\ell(h_j^{p^\prime r}-\langle Z_j \rangle ).\nonumber
				\end{align}\\
				3. Set $\vec{h}^{(p^\prime+1)r }=\vec{h}^{p^\prime r}$. 	
			}	
		}
		Construct the "bias state" of $\{\vec{h}^{p r}\}$, calculate the expectation value $\{E_{p}^r\}$ of $H_\mathrm{C}$.\\
		\Return{The lowest energy $E_p$ in $\{E_p^r\}$ and the corresponding bias state.}
	\end{algorithm}

	There are three major differences from the ab-QAOA, 
 \begin{enumerate}
     \item For the optimization-free ab-QAOA, the output state is constructed from the bias fields $\Vec{h}$ in contrast to $|\psi_\mathrm{f}^{\mathrm{ab}}\rangle$ in the ab-QAOA.
     \item At a fixed level, $\vec{\gamma}$, $\vec{\beta}$ and $\vec{h}$ are updated until convergence in the ab-QAOA. This update is not necessary for the optimization-free ab-QAOA. What is done in level $p^\prime$ is to update the bias fields only once based on those from level $p^\prime-1$. 
    \item For the ab-QAOA, we can directly go to level $p$, while for the optimization-free ab-QAOA, we need to go through all the levels smaller than $p$. These extra levels, labeled by $p'$ with $1 \leq p' \leq p$, are needed to train the bias fields $\vec{h}$.
 \end{enumerate}
    This algorithm is clearly very fast in terms of both classical and quantum resources. The optimization-free ab-QAOA is also different from an iterative QAA procedure~\cite{bias_qaa}, since the Trotter error can not be ignored in the small-level case and the solution is obtained from the bias fields instead of the output state.

    \begin{figure}[htb]
		\centering
		\includegraphics[scale=0.43]{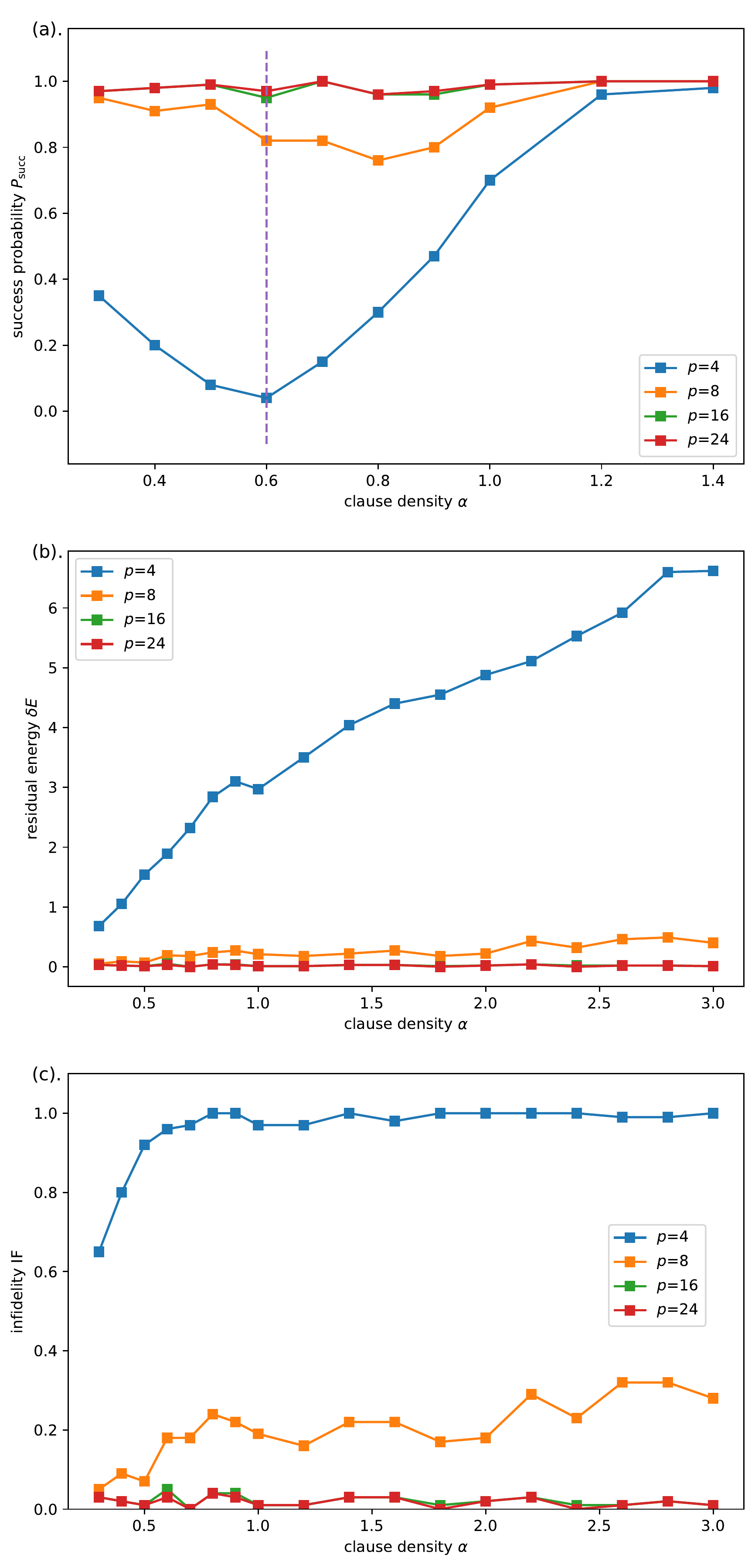}
		\caption{The success probability (a) for $10$-variable $1$-$3$-$\mathrm{SAT}^+$, the residual energy (b) and the infidelity (c) for $10$-variable Max-$1$-$3$-$\mathrm{SAT}^+$ produced by the optimization-free ab-QAOA at levels $4,8,16,24$. The same problem instances as in Figs.~\ref{fig:succ_probability} and~\ref{fig:energy_fidelity} are calculated. The more accurate states with smaller residual energy or infidelity can be obtained with more levels. The numerical results converge at level $16$.  Increasing $R$ or adjusting $\delta t$ can improve the performance further.}\label{fig:energy_fidelity_of}
	\end{figure}
 
    As shown in Fig.~\ref{fig:energy_fidelity_of}, the optimization-free ab-QAOA can also solve the $10$-variable hard-region $1$-$3$-$\mathrm{SAT}^+$ or Max-$1$-$3$-$\mathrm{SAT}^+$ problems effectively and the easy-hard-easy or easy-hard patterns, though present, are not prominent.  In level $16$, perfect solutions can be obtained. However, the result at level $4$ is not as good as the QAOA because only $4$ updates are executed. More levels are needed for training $\vec{h}$ for an accurate state. It seems that although the numerical results converge at level $16$, there is a small fraction of the problems that can not be solved. Increasing $R$ or adjusting $\delta t$ can reduce this fraction. 
	
    Although for the ab-QAOA, the residual energy $\delta E$ and infidelity $\mathrm{IF}$ are about $0$ in level $8$, achieving similar residual energy requires $16$ levels for the optimization-free ab-QAOA.  However, the latter is more efficient in both numerical simulation and realistic experiments since there is no optimization of $\vec{\gamma}$ and $\vec{\beta}$. Similar to the discussion in Sec.~\ref{subsec:qc}, $O(p^2/2)$ quantum gates are required in implementing a $p$-level optimization-free ab-QAOA, as a result, it takes roughly $1/2$ and $1/10$ quantum gates of the original ab-QAOA to solve the hard-region $1$-$3$-$\mathrm{SAT}^+$ or Max-$1$-$3$-$\mathrm{SAT}^+$ problems.
	
	\section{Conclusion}~\label{sec:conclusion}

The age of NISQ devices brings with it many opportunities, but also 
a challenge, of finding a way to exploit quantum advantage for real--world 
problems in a calculation running for a short time, on a relatively small 
number of qubits.
One important class of problems are combinatorial 
optimizations over classical variables, of which SAT problem is  
a canonical example.
Here, the quantum approximate optimization algorithm (QAOA) shows 
great promise~\cite{qaoa_farhi,qaoa_lukin}.
None the less, when the QAOA is applied to the $3$-SAT or Max-$3$-SAT problem, the quantum cost exhibits an easy-hard-easy or easy-hard pattern respectively as the clause density is changed, where the hard-region overhead renders calculations on NISQ devices impractical~\cite{1_k_sat_qaoa,reachabiilty_deficits_sat}.

In this paper, we have explored the possibility of designing a QAOA--like algorithm  
which is capable of solving the hard combinatorial optimization problems on NISQ devices.
We did so in the context of the SAT problem, comparing the performance of the QAOA with a recently--introduced variant, 
 the adaptive-bias quantum approximate optimization algorithm (ab-QAOA)~\cite{ab_qaoa}.
 As specific examples we considered $1$-$3$-$\mathrm{SAT}^+$ or Max-$1$-$3$-$\mathrm{SAT}^+$ 
 problems, for up to 14 variables.
 We find that the easy-hard-easy or the easy-hard pattern in the ab-QAOA is nearly absent and the ab-QAOA offers a considerable reduction in the resources required to solve these problems in the hard region.
 In particular, the ab-QAOA can solve the $10$-variable $1$-$3$-$\mathrm{SAT}^+$ and Max-$1$-$3$-$\mathrm{SAT}^+$ problems at much lower circuit depth, finding solutions with nearly perfect fidelity in levels $4$ and $8$ respectively, while for the QAOA, level $24$ or sometimes even level $64$ is far from enough.
 
 By estimating the levels that QAOA and ab-QAOA need to achieve the same accuracy for the $10$-variable $1$-$3$-$\mathrm{SAT}^+$ or Max-$1$-$3$-$\mathrm{SAT}^+$ problems, we find that a $50$-fold or respectively, a $90$-fold speedup is realized in the ab-QAOA,  
implying that the ab-QAOA greatly improves performance for the hard-region $3$-SAT problems and hard-region Max-$3$-SAT problems. 
For the hard-region 14-variable problems, a level-8 ab-QAOA can obtain the solution with fidelity about $0.85$ in contrast to $0.1$ in the level-8 QAOA. 
Both the reduction in the circuit depth required to achieve a given accuracy, and the nearly absent dramatic change 
in the resources required as a function of clause density, suggest that ab--QAOA is more likely than QAOA to demonstrate the quantum advantages for this class of problems, on coming quantum devices.
%

By itself, this work does not give any direct evidence for a quantum speedup over classical algorithms.  
Indeed, the systems simulated in this paper are sufficiently small that classical algorithms can very quickly 
solve the problems even in the hard regions.  
However, the present work does shed some light on the type of problem where such a speedup might be found in the future. 
We have found a set of problems with a hardness parameter (clause density) which seems to be difficult for both 
a non-adaptive quantum algorithm and all classical algorithms, for a certain range of the parameter. 
In contrast, an adaptive quantum algorithm does not have the same dependence on the parameter.

	The efficiency of the ab-QAOA is connected to the special nature of the solutions of Ising-model optimization problems. These solution states can be encoded into the mixing Hamiltonian by the use of bias fields. This leads to lower entanglement and more localization during the evolution, greatly enhancing the adiabaticity. And this speaks to a problem at the heart of quantum computation in the NISQ era: it is entanglement that makes the quantum computation different from its classical counterpart, but too much entanglement in an algorithm induces the barren plateau phenomenon~\cite{barren_plateaus_ent,barren_plateaus_ent2,barren_plateaus_avoid}.    
  
    From this point of view, the ab-QAOA appears to be in the ''Goldilocks zone'' for entanglement, providing a concrete algorithm which restricts  entanglement in the course of the optimization, without losing computational power. The benefit of such an approach has already been discussed~\cite{barren_plateaus_ent2,barren_plateaus_avoid,ADAPTqaoa_ent}, and the results in this article strongly suggest that localizing the state on the solution set, as directly measured by the participation ratio, is a good way to achieve a compromise between too much and too little entanglement.  The ab-QAOA shows that for a certain very important class of problems, this can be achieved without increasing the computational cost. 

    A much simpler optimization-free ab-QAOA is also proposed. Training the bias fields parameters within the ab-QAOA in the increasing levels without any need of optimizing $\vec{\gamma}$ and $\vec{\beta}$ can find the solutions of the $10$-variable $1$-$3$-$\mathrm{SAT}^+$ or Max-$1$-$3$-$\mathrm{SAT}^+$ problems in about level $16$. It appears to be more powerful than the QAOA for these hard problem instances with only $1/2$ or $1/10$ quantum gates of the original ab-QAOA. The ab-QAOA or the optimization-free ab-QAOA can bring a quantum advantage closer in real-world applications.
 
	An important open issue is the initialization of the bias field parameters $\vec{h}$, which we randomly generate in the current work. This was clearly sufficient to solve the instances we considered, which had a problem size $n=10$. It is necessary to understand how the accuracy of the initial guess of the $\vec{h}$  scales with $n$ to obtain the same infidelity, but this is difficult without very large computational resources.  A non-random approach would be to run a classical heuristic to get the initial value of $\vec{h}$. This is similar to warm-start approaches to quantum optimization \cite{qaoa_warmstart,cain2022qaoa}. For the purposes of this paper, this method would obscure the pure effect of adding bias fields to the QAOA, but could be a subject of future investigation. 

    The source code and the raw data of the problem definition can be found in~\cite{codes}.
	
	\begin{acknowledgments} 
		Yunlong Yu and Xiang-Bin Wang acknowledge National Natural Science Foundation of China Grants No. 11974204 and No. 12174215. Nic Shannon acknowledges the support of the Theory of Quantum Matter Unit, Okinawa Institute of Science and Technology Graduate University (OIST).  
	\end{acknowledgments}
	
	\appendix
  
	\section{Modified Fourier strategy}\label{sec:modified_lukin}
\begin{figure}[H]
		\centering
		\includegraphics[scale=0.43]{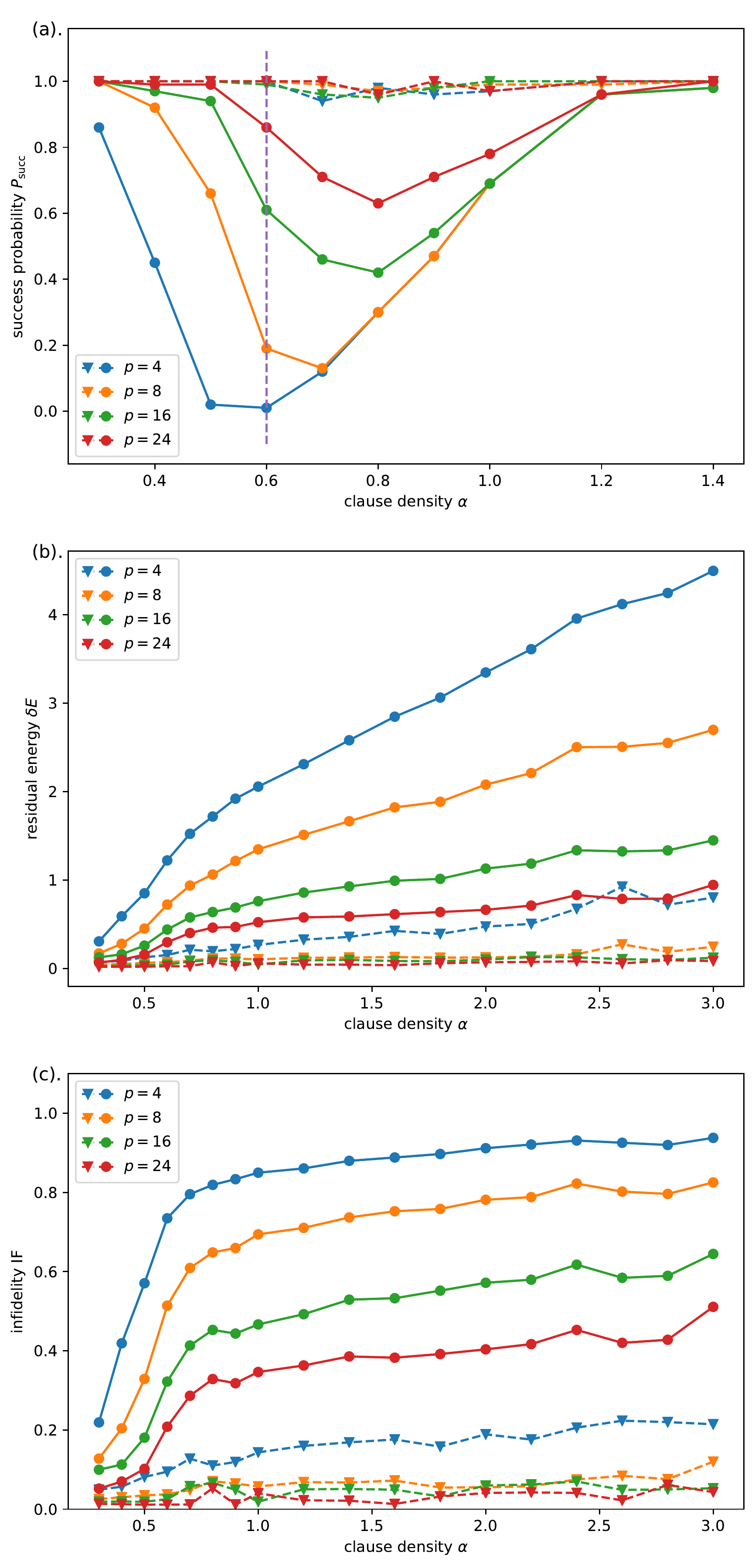}
		\caption{Comparison of the success probability (a), residual energy (b) or the infidelity (c) of QAOA (solid lines) to that of the ab-QAOA (dashed lines) for level $p=4,8,16,24$ . The same problem instances as Figs. \ref{fig:succ_probability} and \ref{fig:energy_fidelity} are calculated. Both the QAOA and ab-QAOA are initialized through the modified Fourier strategy.}\label{fig:energy_fidelity2}
	\end{figure}

	In Algorithm~\ref{alg:modified_fourier} and Fig~\ref{fig:energy_fidelity2}, we give details of our modification of the Fourier strategy of Ref.~\cite{qaoa_lukin} as applied to the Max-$1$-$3$-$\mathrm{SAT}^+$ problems. It was invented for the QAOA and has been adapted for the ab-QAOA in Ref.~\cite{ab_qaoa}. Since the QAOA is the $h_j\rightarrow 0$ and $\ell\rightarrow 0$ limitation of the ab-QAOA, it is sufficient to take the ab-QAOA as an example.

    In the original Fourier strategy, for the state $|\psi_\mathrm{f}^\mathrm{ab}(\vec{\gamma},\vec{\beta},\vec{h})\rangle$ in fixed level $p^\prime$, the Fourier transforms of $\vec{\gamma}$ and $\vec{\beta}$ are optimized instead of $\vec{\gamma}$ and $\vec{\beta}$ themselves. $R$ initial points are optimized in parallel and the optimized points with the best energy are chosen as the output points and energies in this level. How to choose the $R$ initial points in level $p^\prime+1$ depends on the output point in level $p^\prime$ starting from a randomly initialized point in level $1$. 
    
    As shown in ~\cite{qaoa_lukin,ab_qaoa}, the Fourier strategy for MaxCut avoids some points that are not favorable, perhaps corresponding to local minima.  In the SAT problems we found that this strategy is  not efficient at higher levels and in fact the straightforward optimization of $\vec{\gamma}$ and $\vec{\beta}$ was generally superior. So we propose a modified Fourier strategy in Algorithm~\ref{alg:modified_fourier} to circumvent these two issues. The main idea is that the $R$ initial points in level $p^\prime$ can be constructed from the output point in any level smaller than $p^\prime$ instead of only $p^\prime-1$. The codes are available in~\cite{codes}.

	 In Algorithm~\ref{alg:modified_fourier}, the superscripts $t$ means the $t^\mathrm{th}$ level in the level list, the superscript $r$ means the $r^\mathrm{th}$ sample in $R$ samples and the superscript $\mathrm{B}$ means the best point with the lowest energy among $R$ energies. The random vector $\mathrm{Ran}(\vec{u})$, is the same length as $\vec{u}$, and its $k^\mathrm{th}$ component is a normally-distributed number with mean $0$ and variance $u_k^2$ multiplied by a factor $\xi$. In our calculations, the level list is $\{1,2,4,8,16,24\}$ and $\xi=0.6$. Note that this modified strategy also applies to the QAOA except that $h_j=0$ and $\ell=0$.

    \begin{algorithm}
		\SetKwInOut{Return}{Return}
		\label{alg:modified_fourier}
		\caption{Modified Fourier strategy for level-$p$ ab-QAOA}
		\KwIn{level list $\{p_1,p_2,\cdots,p_T\}$ in increasing order with $p_T=p$, sample number $R$} 
		\KwOut{optimized energy list $\{E_{p_t}\}$ from $t=1$ to $T$. }
		\For{$t=1$ \KwTo $T$}{
			Randomly generate $R$ bias fields parameters $\{\vec{h}^{tr}\}$ with each component to be $1$ or $-1$.\\
			\eIf{$t\,\mathrm{is}\, 1$}{
				1. Randomly generate $R$ points $\{(\vec{\gamma}^{1r},\vec{\beta}^{1r})\}$.\\ 
				2. Optimize over these $R$ points in parallel and find the optimal point $(\vec{\gamma}^{1\mathrm{B}},\vec{\beta}^{1\mathrm{B}},\vec{h}^{1\mathrm{B}})$ with the lowest energy $E_{p_1}$  .
			}
			{ 1. The $R$ initial $\{(\vec{\gamma}^{tr},\vec{\beta}^{tr})\}$ are generated according to the recipe below,\\
				\eIf{$r\,\mathrm{is}\, 1$}{
					The first $p_{t-1}$	components of $\vec{\gamma}^{tr}$ or $\vec{\beta}^{tr}$ are exactly $\vec{\gamma}^{(t-1)\mathrm{B}}$ or $\vec{\beta}^{(t-1)\mathrm{B}}$. The remaining components are $0$.
				}
				{
					The first $p_{t-1}$	components of $\vec{\gamma}^{tr}$ or $\vec{\beta}^{tr}$ are $\vec{\gamma}^{(t-1)\mathrm{B}}$ or $\vec{\beta}^{(t-1)\mathrm{B}}$ added with random vectors $\mathrm{Ran}(\vec{\gamma}^{(t-1)\mathrm{B}})$ or $\mathrm{Ran}(\vec{\beta}^{(t-1)\mathrm{B}})$. The remaining components are $0$.
				}
				2. Optimize over these $R$ points in parallel and find the optimal point $(\vec{\gamma}^{t\mathrm{B}},\vec{\beta}^{t\mathrm{B}},\vec{h}^{t\mathrm{B}})$ with the lowest energy $E_{p_t}$.			
			}
		}
		\Return{The optimized energy list $\{E_{p_t}\}$.}
	\end{algorithm}

    In Fig.~\ref{fig:energy_fidelity2}, the same problem instances as those in  Sec.~\ref{sec:result_energy} and Sec.~\ref{sec:result_energy} are calculated with only the initialization method different. There is little difference compared with Figs.~\ref{fig:succ_probability} and~\ref{fig:energy_fidelity} except that the modified Fourier strategy is better for the high-level QAOA. The reachability deficits of the QAOA are only slightly mitigated with the new initialization method but more substantially mitigated in the ab-QAOA at level $8$.
	
    \section{Reachability deficits beyond $\alpha=3$ and $n=10$ }\label{sec:ef_4}
    
\begin{figure}[H]
		\centering
        \includegraphics[scale=0.41]{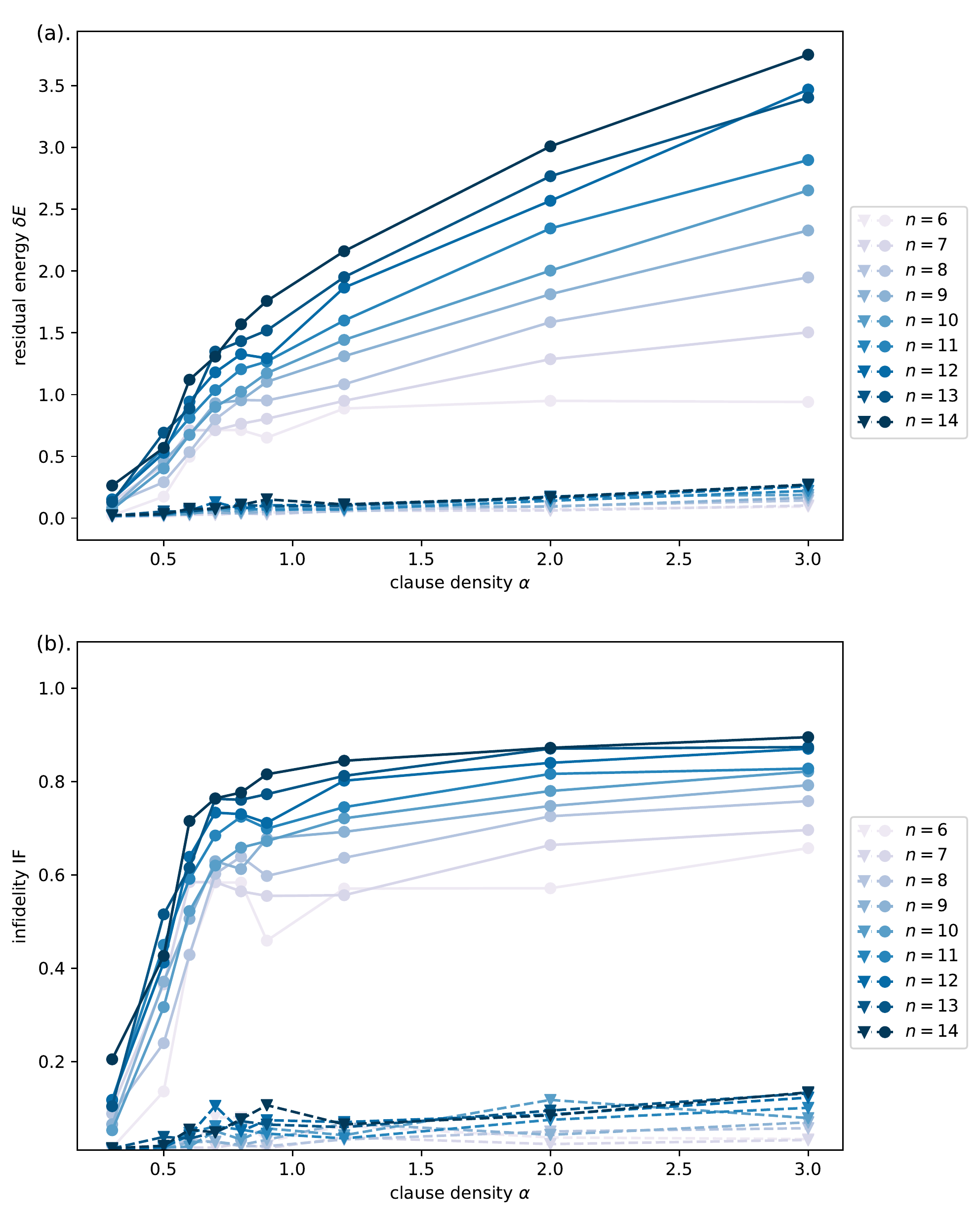}
		\caption{Comparison of the residual energy (a) and infidelity (b) of QAOA (solid lines) to that of the ab-QAOA (dashed lines) for the Max-1-3-SAT$^+$ problems. Results are given for levels $p=8$ as a function of the clause density $\alpha$ and $n$.  Each point is an average over $100$ random instances. }\label{fig:energy_fidelity4}
	\end{figure}

\begin{figure}[H]
		\centering
		
        \includegraphics[scale=0.43]{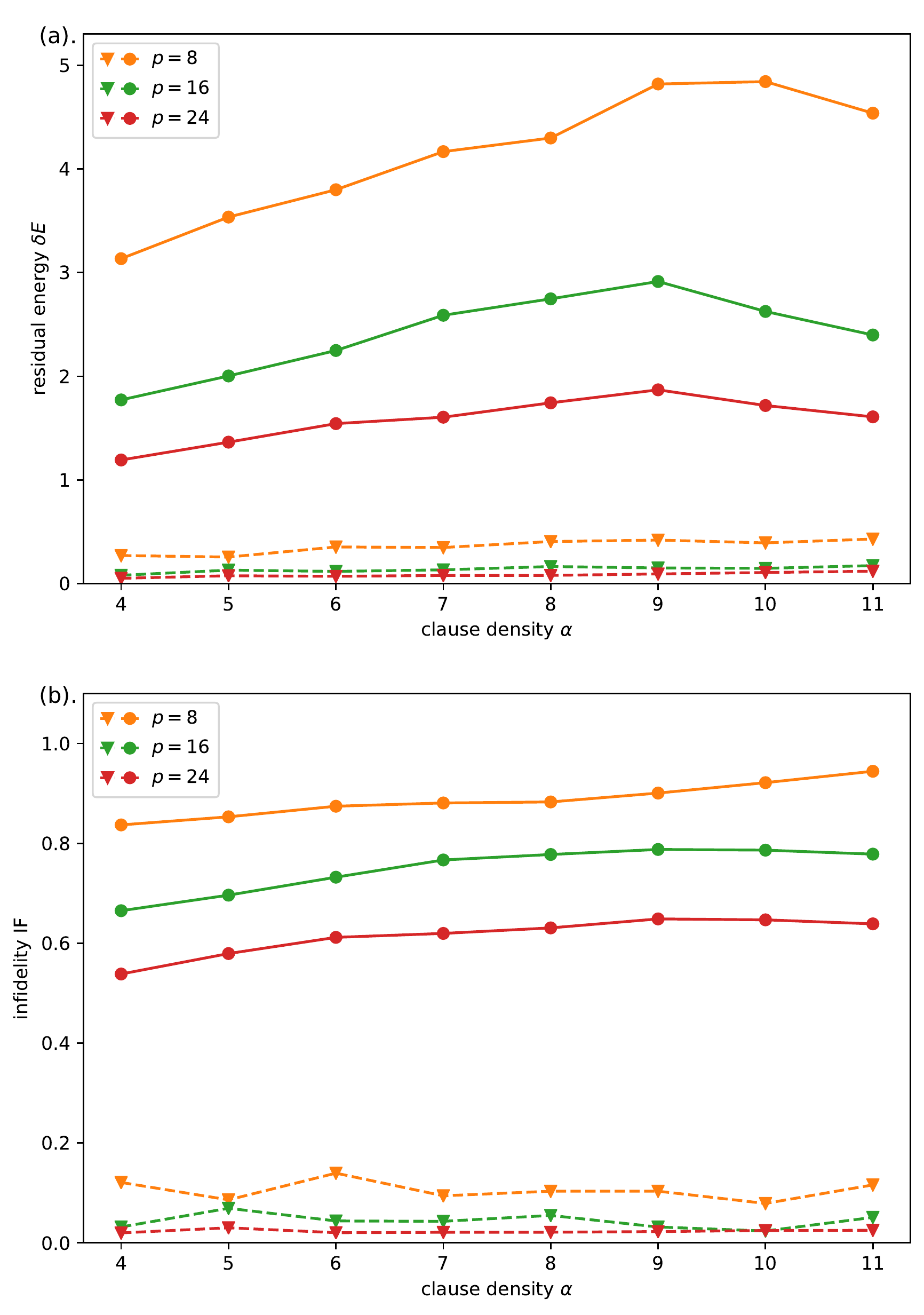}
		\caption{Comparison of the residual energy (a) and infidelity (b)  of QAOA (solid lines) to that of the ab-QAOA (dashed lines) for the $10$-variable Max-1-3-SAT$^+$ problems. Results are given for levels $p=8,16,24$ as a function of the clause density $\alpha$, where the increment of $\alpha$ is $1$ from $\alpha=4$ to $11$. Each point is an average over $100$ random instances. The Modified Trotterized quantum annealing method is applied to the QAOA and ab-QAOA with $\delta t=0.2$. The ab-QAOA can solve the problems effectively at level $8$. }\label{fig:energy_fidelity3}
	\end{figure}

     In this Appendix, the results for the residual energy and infidelity for $n\leq 14$ with selected $\alpha$  and for $n=10$ with $\alpha\geq4$ are given in Figs.~\ref{fig:energy_fidelity4} and~\ref{fig:energy_fidelity3}. 

    As shown in Fig.~\ref{fig:energy_fidelity4}, as the problem hardness increasing with increasing $n$, the residual energy and infidelity also increase. This implies that more levels are needed to achieve a given accuracy, which is consistent with Fig.~\ref{fig:cost}. Note that when $\alpha=3$ and $n=14$, the infidelity of the QAOA is about $0.1$ in contrast to $0.85$ in the ab-QAOA, which shows the superiority in the ab-QAOA.
     
     For the $10$-variable $1$-$3$-$\mathrm{SAT}^+$ and Max-$1$-$3$-$\mathrm{SAT}^+$ problems, the maximal clause density is $12$. The maximal clause density in the calculation is $\alpha=11$. The reachability deficits will be more evident with $\alpha$ approaching $11$ where a small number of the levels is not sufficient. So as shown in Fig.~\ref{fig:energy_fidelity3}, only the results in $p=8,16,24$ are presented. The reachability deficits are still evident for the QAOA, while
    they are less evident in the ab-QAOA even at level $8$. As analyzed in Sec.~\ref{subsec:qc}, the cost Hamiltonian tends to the identity matrix and the gap between the ground state and the first excited state is amplified when $\alpha$ approaches $11$, leading to the improved performances of the QAOA.

	\section{Entanglement entropy, participation ration and annealing entropy for $\alpha=0.8$}\label{sec:res_a8}
	
\begin{figure}[htb]
		\centering
		\includegraphics[scale=0.43]{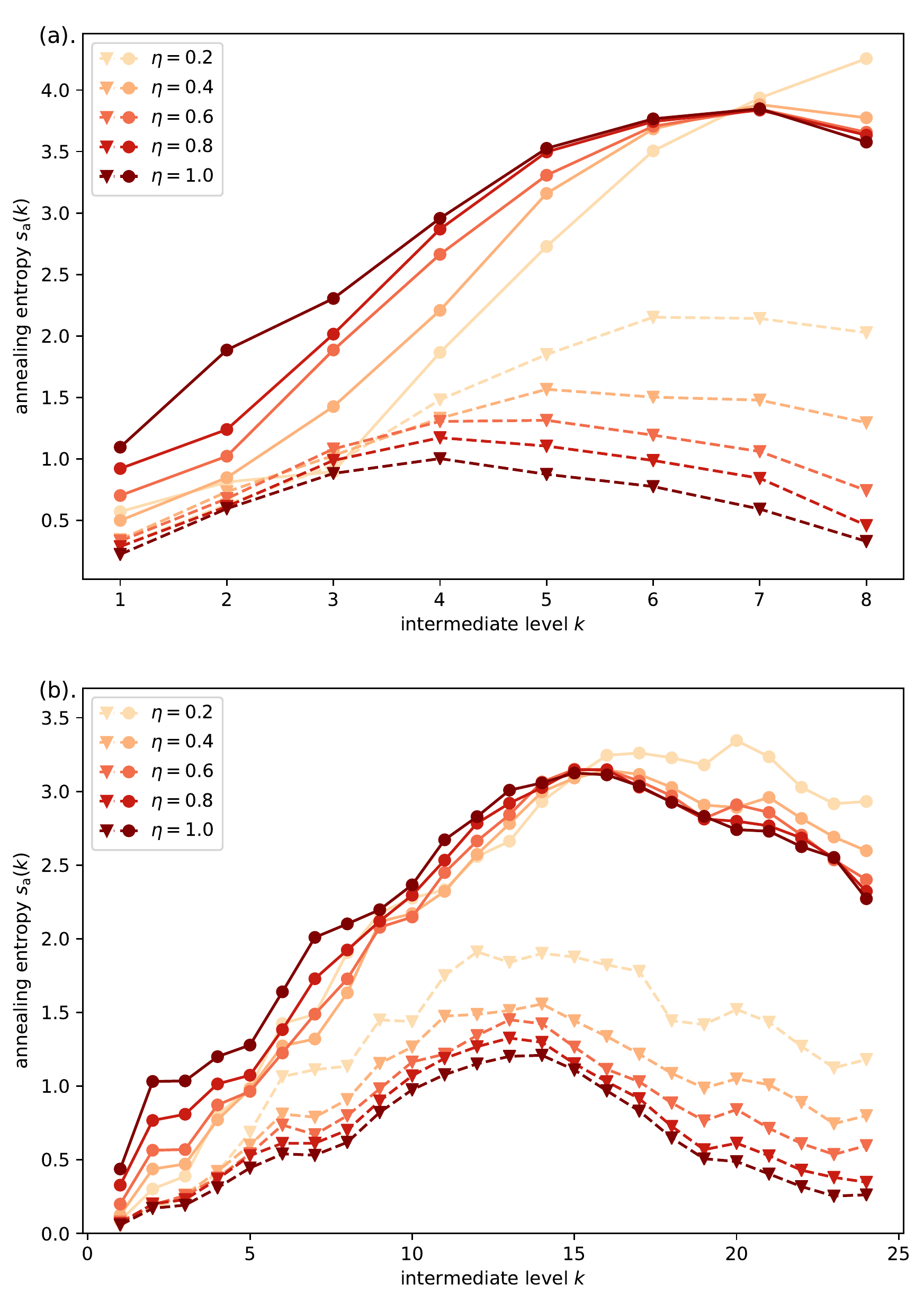}
		
		\caption{The dynamical evolution of the annealing entropy $s_\mathrm{a}(k)$ of the QAOA (solid lines) and the ab-QAOA (dashed lines) with level $p=8$ (a) and $p=24$ (b). The problem instances are those with $\alpha=0.8$ in Fig.~\ref{fig:energy_fidelity}. The ab-QAOA is more adiabatic because of the much smaller annealing entropy.}\label{fig:annealing2}
	\end{figure}

\begin{figure*}[htb]
		\centering
		\includegraphics[scale=0.43]{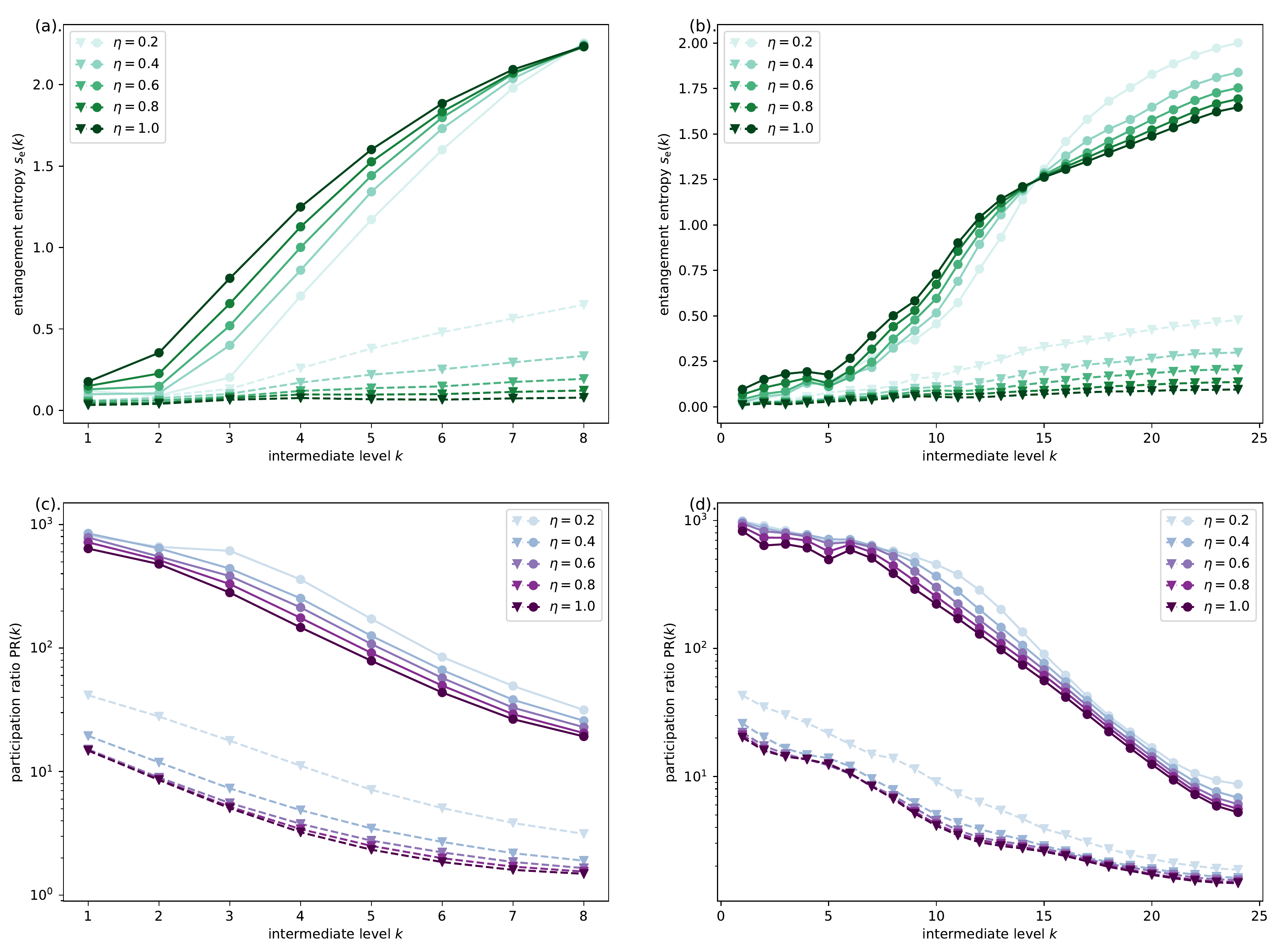}
		\caption{The dynamical evolution of the intermediate entanglement entropy $s_\mathrm{e}(k)$ (a, b) and the intermediate inverse participation ratio (c, d) of the QAOA (solid lines) and the ab-QAOA (dashed lines) with level $p=8$ (a, c) and $p=24$ (b, d). The problem instances are those with $\alpha=0.8$ in Fig.~\ref{fig:energy_fidelity}. The parameter $\eta$ is the dynamical parameter in the optimization. The ab-QAOA is less entangled and more localized because of the localized bias fields in the mixing Hamiltonian.}\label{fig:entangle_pr2}
	\end{figure*}

	In this Appendix, we give the additional numerical results for the hard-region $1$-$3$-$\mathrm{SAT}^+$ problems with $\alpha=0.8$. The annealing entropy is shown in Fig.~\ref{fig:annealing2}. The entanglement entropy and participation ratio are shown in Fig.~\ref{fig:entangle_pr2}. 

    \newpage
	\bibliography{ref}

\end{document}